\newif\ifisit
\isitfalse

\newif\ifproofs
\ifisit
\proofsfalse
\else
\proofstrue
\fi
\documentclass[journal,12pt,draftclsnofoot,onecolumn]{IEEEtran}
\IEEEoverridecommandlockouts
\usepackage{amssymb}
\usepackage{amsmath}
\usepackage{amsbsy}
\usepackage{graphics,graphicx}
\usepackage[us]{datetime}
\usepackage{verbatim}
\usepackage{bm}
\usepackage{rotating}
\usepackage{accents}
\newtheorem{theorem}{Theorem}
\newtheorem{proposition}{Proposition}

\newtheorem{lemma}{Lemma}
\newtheorem{corollary}{Corollary}
\newenvironment{proof}{\begin{IEEEproof}}{\end{IEEEproof}}
\newtheorem{@example}{Example}
\newenvironment{example}{\begin{@example}\upshape}{\end{@example}}

\renewcommand{\emptyset}{\phi}
\newcommand{\blds}{\mathbf{s}}
\newcommand{\bldk}{\mathbf{k}}
\newcommand{\bldkbar}{\bar{\mathbf{k}}}

\newcommand{\bldz}{\mathbf{z}}
\newcommand{\bldZ}{\mathbf{Z}}
\newcommand{\ord}{\mathrm{ord}}
\newcommand{\alp}{A}
\newcommand{\alpF}{\mathcal{A}}
\newcommand{\alphabet}{\mathcal{A}}
\newcommand{\states}{\mathcal{S}}

\newcommand{\nocc}[1]{N_{#1}}
\newcommand{\nonempty}{\neq \emptyset}
\newcommand{\dominant}{dominant}
\newcommand{\totdominant}{totally dominant}
\newcommand{\domfree}{domination-free}

\newcommand{\bigstrlen}[1]{\bigl|#1\bigr|}

\newcommand{\alphseq}[1][\partn]{\mathbf{A}_{#1}}

\newcommand{\swltr}{\mathrm{w}}
\newcommand{\pswitch}{P_\swltr}
\newcommand{\fswitch}{f_\swltr}
\newcommand{\pswitchp}{P'_\swltr}
\newcommand{\pswitchpp}{P''_\swltr}

\newcommand{\kswitch}{k_\swltr}
\newcommand{\intlv}{\mathcal{I}}

\newcommand{\partn}{\Pi}

\newcommand{\partnp}{\partn'}
\newcommand{\partnpp}{\partn''}
\newcommand{\partnbar}{\bar{\partn}}

\newcommand{\kparm}{\bldk}
\newcommand{\kvec}{\bldk}

\newcommand{\sti}[1]{{s^{(#1)}}}
\newcommand{\stip}[1]{{{s'}^{\raisebox{-6pt}{$\scriptstyle(#1)$}}}}
\newcommand{\ISMP}{IMP}

\newcommand{\IPi}[1][\partn]{\intlv_{#1}}
\newcommand{\IPip}{\IPi[\partnp]}

\newcommand{\cost}{C}

\newcommand{\FPk}[2][\partn]{\mathcal{F}_{#1,#2}}
\newcommand{\FPkp}[1][\partn']{\FPk[#1]{\kparm'}}
\newcommand{\FP}[1][\partn]{\FPk[#1]{\kparm}}

\newcommand{\PFP}[1][\partn]{P_{#1,\kparm}}
\newcommand{\PF}[1][F]{P_{#1}}

\newcommand{\canon}[2][P]{(#2)^\ast_{#1}}

\newcommand{\VVV}{\mathcal{V}}
\newcommand{\PPP}{\mathcal{P}}
\newcommand{\VFP}[1][\FP]{\VVV({#1})}
\newcommand{\VFPI}[1][\FP]{\VVV_{\intlv}({#1})}
\newcommand{\VFPb}[1][\FP]{\overline{\VVV}({#1})}
\newcommand{\VFPIb}[1][\FP]{\overline{\VVV}_{\intlv}({#1})}
\newcommand{\Vclosed}{\overline{V}}
\newcommand{\mcPFPI}[1][\FP]{\PPP_{\intlv}({#1})}
\newcommand{\cref}{+}
\newcommand{\dpartn}{\mathcal{K}}
\newcommand{\dpartni}{\kappa}
\newcommand{\dpartnib}{\bar{\dpartni}}

\newcommand{\half}{\frac{1}{2}}
\newcommand{\proj}[2]{#1[#2]}

\newcommand{\Pbadp}[1][\partnp,\bldk']{\mathcal{P}_{#1}}
\newcommand{\Perr}{P_{\text{err}}}

\newcommand{\defined}{\stackrel{\Delta}{=}}

\newcommand{\refF}[1][F]{{#1}^{+}}

\newcommand{\hhk}[1][k]{\hat{H}_{#1}}
\newcommand{\hhF}[1][\partn,\bldk]{\hat{H}_{#1}}
\newcommand{\hpF}[1][\partn,\bldk]{\hat{P}_{#1}}
\newcommand{\hhFp}{\hhF[{\partnp,\bldk'}]}
\newcommand{\hpFp}{\hpF[{\partnp,\bldk'}]}

\newcommand{\kcost}[1][k]{C_{#1}}

\newcommand{\chubchik}[1]{\accentset{\approx}{#1}}

\newlength{\dhatheight}
\newcommand{\doublehat}[1]{%
\settoheight{\dhatheight}{\ensuremath{\hat{#1}}}%
\addtolength{\dhatheight}{-0.25ex}%
\hat{\vphantom{\rule{1pt}{\dhatheight}}%
\smash{\hat{#1}}}}
\newcommand{\kumkum}[1]{\doublehat{#1}}

\newcommand{\pml}{\hat{P}}
\newcommand{\pmlk}[1][k]{\pml_{#1}}

\newcommand{\pmlF}[1][F]{\pml^{\ast}_{#1}}
\newcommand{\Ponej}[1][j]{P_1^{(#1)}}

\newcommand{\onem}{\{1,2,\ldots,m\}}
\newcommand{\twom}{\{2,3,\ldots,m\}}
\newcommand{\onemw}{\{1,2,\ldots,m,\swltr\}}
\newcommand{\onetwo}{\{1,2\}}
\newcommand{\sspace}{\bm{\Pi}}
\newcommand{\badset}{B_{\partnp,\bldk'}}
\newcommand{\uut}{u^{(t)}}
\newcommand{\uus}{u^{(s)}}

\newcommand{\dom}{\sqsupset}
\newcommand{\ndom}{\not\dom}
\newcommand{\mdom}{\raisebox{0.3ex}{$\scriptstyle\sqsupset\sqsubset$}}
\newcommand{\psimapfull}[1][\partn,\partnp]{\Psi_{#1}}
\newcommand{\psimap}[1][{}]{\Psi_{#1}}
\newcommand{\LLL}{M}
\newcommand{\MMM}{M_2}
\newcommand{\bigdiv}[2]{D\bigl(#1\bigl|\bigl|\bigr.\bigr.#2\bigr)}
\newcommand{\neighbor}{\mathcal{N}}

\newcommand{\Remarks}{\noindent\textbf{Remarks. }\ }

\begin{document}

% for some reason the earlier definition doesn't stick
\renewenvironment{proof}{\begin{IEEEproof}}{\end{IEEEproof}}

\title{Deinterleaving Finite Memory Processes via Penalized Maximum Likelihood}
% author names and affiliations
% use a multiple column layout for up to three different
% affiliations
\author{
\normalsize
\IEEEauthorblockN{Gadiel Seroussi\\[-1ex]}
\IEEEauthorblockA{Hewlett-Packard Laboratories\\[-1.5ex]
Palo Alto, CA, USA \\[-1.5ex]
gseroussi@ieee.org\\} %
\and \IEEEauthorblockN{Wojciech Szpankowski\\[-1ex]}
\IEEEauthorblockA{ %Department of Computer Science\\
Purdue University,\\[-1.5ex]
West Lafayette, IN, USA\\[-1.5ex]
spa@cs.purdue.edu\\} %
\and \IEEEauthorblockN{Marcelo J.\ Weinberger\\[-1ex]}
\IEEEauthorblockA{Hewlett-Packard Laboratories\\[-1.5ex]
Palo Alto, CA, USA \\[-1.5ex]
marcelo.weinberger@hp.com}
\thanks{W. Szpankowski's work was partially done while visiting HP Labs, Palo
Alto, CA, and also supported by NSF Science and Technology Center
Grants CCF-0939370 and CCF-0830140.}}

\maketitle
\thispagestyle{empty} % to unnumber title page

\begin{abstract}
We study the problem of deinterleaving a set of finite-memory (Markov)
processes over disjoint finite alphabets, which have been randomly
interleaved by a finite-memory switch. The deinterleaver has access to
a sample of the resulting interleaved process, but no knowledge of the
number or structure of the component Markov processes, or of the
switch. We study conditions for uniqueness of the interleaved
representation of a process, showing that certain switch
configurations, as well as memoryless component processes, can cause
ambiguities in the representation. We show that a deinterleaving scheme
based on minimizing a penalized maximum-likelihood cost function is
strongly consistent, in the sense of reconstructing, almost surely as
the observed sequence length tends to infinity, a set of component and
switch Markov processes compatible with the original interleaved
process. Furthermore, under certain conditions on the structure of the
switch (including the special case of a memoryless switch), we show
that the scheme recovers \emph{all} possible interleaved
representations of the original process. Experimental results are
presented demonstrating that the proposed scheme performs well in
practice, even for relatively short input samples.
\end{abstract}

\section{Introduction}
Problems in applications such as data mining, computer forensics,
finance, and genomics, often require the identification of streams of
data from different sources, which may be intermingled or hidden
(sometimes purposely) among other unrelated streams, in large
interleaved record files. In this haystack of records can lie buried
valuable information whose extraction would be easier if we were able
to separate the contributing streams. The deinterleaving problem
studied in this paper is motivated by these applications (more detailed
accounts of which can be found, for example,
in~\cite{colt04,Landwehr08,Gillblad_et_al'2009}).

In our setting, the data streams, as well as the interleaving agent,
will be modeled as sequences generated by discrete-time random
processes over finite alphabets. Specifically, let
$\alp_1,\,\alp_2,\,\ldots\,,\,\alp_m$ be finite, nonempty, disjoint
alphabets, let $\alphabet = A_1 \cup A_2
\cup \cdots A_m$, and $\partn=\{\alp_1,\alp_2,\ldots,\alp_m\}$. We
refer to the $A_i$ as
\emph{subalphabets}, and to $\partn$ as a \emph{partition}, of
$\alphabet$. Consider $m$ independent, \emph{component random
processes} $P_1,P_2,\ldots,P_m$, defined, respectively,  over
$\alp_1,\,\alp_2,\,\ldots\,,\,\alp_m$, and a random \emph{switch
process} $\pswitch$ over the alphabet $\partn$, independent of
the component processes. %
The \emph{interleaved process}
$P\defined\IPi(P_1,P_2,\ldots,P_{m};\pswitch)$ is generated as follows:
At each time instant, a subalphabet $A_i\in\partn$ is selected
according to $\pswitch$, and the next output sample for $P$ is selected
from $A_i$ according to the corresponding process $P_i$ (we say,
loosely, that the switch ``selects'' $P_i$ at that instant). The
component processes $P_i$ are idle when not selected, i.e., if $P_i$ is
selected at time $t$, and next selected at time $t+T$, then the samples
emitted by $P$ at times $t$ and $t+T$ are
\emph{consecutive} emissions from $P_i$, regardless of the length of
the intervening interval~$T$.

Given a sample $z^n$ from $P$, and without prior knowledge of the
number or the composition of the subalphabets $A_i$, the
\emph{deinterleaving} problem of interest is to reconstruct the original sequences
emitted by the component processes, and the sequence of switch
selections.

So far, we have made two basic assumptions on the structure of the
interleaved system: the independence of the component and switch
processes, and the disjointness of the subalphabets. The latter
assumption implies that, given an interleaved input stream, identifying
the partition $\partn$ is equivalent to identifying the component
substreams and the sequence of switch selections. Thus, identifying the
partition $\partn$ is sufficient to solve the deinterleaving problem.
Identifying the substreams when the subalphabets are not disjoint is
also a problem of interest, but it appears more
challenging~\cite{colt04}, and is outside the scope of this paper. Even
with these assumptions, it is clear that without further restrictions
on the component and switch processes, the problem defined would be
either ill-posed or trivial, since two obvious hypotheses would always
be available: the interleaved process $P$ could be interpreted as
having a single component $P_1=P$, or  as an interleaving of constant
processes over singleton alphabets interleaved by a switch $\pswitch$
essentially identical to $P$. Therefore, for the problem to be
meaningful, some additional constraints must be posed on the structure
of the component and switch processes. In this paper, we study the case
where the components and switch are ergodic
\emph{finite memory} (Markov) processes, i.e., for each $i\in\onemw$,
there is an integer $k_i\ge 0$ such that for any sufficiently long
sequence $u^t$ over the appropriate alphabet, we have $P_i(u_t|u^{t-1})
= P_i(u_t|u_{t-k_i}^{t-1})$. We assume no knowledge or bound on the
process orders $k_i$, and refer to $P$ in this case as an
\emph{interleaved Markov process} (IMP). Except for some
degenerate cases (e.g., when all the component processes are
memoryless), the IMP $P$ is generally \emph{not} a finite memory
process, since the interval between consecutive selections of a
component process is unbounded. Hence, in general, the two obvious
hypotheses mentioned above are not available, and the deinterleaving
problem for IMPs is well-posed, non-trivial, and, as we shall show,
solvable.

When $P=\IPi(P_1,P_2,\ldots,P_{m};\pswitch)$ for finite memory
processes $P_1,P_2,\ldots,P_m,\pswitch$, we say that $\partn$ is
\emph{compatible} with $P$, and refer to
$\IPi(P_1,P_2,\ldots,P_{m};\pswitch)$ also
 as an \emph{IMP representation} of $P$. Notice that, given
an IMP $P$, any partition $\partn'$ of $\alphabet$ induces a set of
deinterleaved component and switch processes. In general, however, if
$\partn'$ is the ``wrong'' partition (i.e., it is incompatible with
$P$), then either some of the induced sub-processes $P'_i$ or
$\pswitchp$ will not be of finite order, or some of the independence
assumptions will be violated. There could, however, be more than one
``right'' partition: IMP representations need not be unique, and we may
have partitions $\partn{\ne}\partnp$ such that both $\partn$ and
$\partnp$ are compatible with $P$. We refer to this situation as an
\emph{ambiguity}
in the IMP representation of $P$.\footnote{%
Notice that since $P$ and $\partn$ uniquely determine the component and
switch processes, two different IMP representations of the same process
$P$ \emph{must} be based on different partitions.}

In this paper, we study IMP ambiguities, derive conditions for
uniqueness of IMP representations, and present a deinterleaving scheme
that identifies, eventually almost surely, an IMP representation of the
observed process. Under certain conditions, including all the cases
where the switch is memoryless, the scheme will identify
\emph{all} IMP representations of the process. The solution is based on
finding a partition $\partn$ of $\alp$ and  an \emph{order vector}
$\bldk=(k_1,k_2,\ldots,k_m,k_\swltr)$ that minimize a
\emph{penalized maximum-likelihood} (penalized ML) cost function of the
form $\cost_{\partn,\bldk}(z^n) =n\hat{H}_{\partn,\bldk}(z^n) +\beta
\dpartni
\log n$, where $\hat{H}_{\partn,\bldk}(z^n)$ is the empirical entropy of the observed
sequence $z^n$ under an
\ISMP\ model induced by $\partn$ and $\bldk$, $\dpartni$ is the total number of free
statistical parameters in the model, and $\beta$ is a nonnegative
constant. Penalized ML estimators of Markov process order are well
known (cf.~\cite{Schwartz78,Rissanen78,CsiszarShields00}). Here, we use
them to estimate the original partition $\partn$, and also the Markov
order of the processes $P_i$ and the switch $\pswitch$.

The deinterleaving problem for the special case where all processes
involved are of order at most one has been previously studied
in~\cite{colt04}, where an approach was proposed that could identify an
\ISMP\ representation of $P$ with high probability as $n{\to}\infty$
(the approach as described cannot identify multiple solutions when they
exist; instead, all cases leading to possible ambiguities are excluded
using rather coarse conditions). The idea is to run a greedy sequence
of tests, checking equalities and inequalities between various event
probabilities (e.g., $P(ab){\ne}P(a)P(b),\,P(abc) =
P(a)P(b)P(c),\,a,b,c \in \alphabet$), and permanently clustering
symbols into subalphabets sequentially, according to the test results
(sequentiality here is with respect to the alphabet processing, not the
input sequence, which has to be read in full before clustering begins).
Empirical distributions are used as proxies for the true ones. Clearly,
equalities between probabilities translate only to ``approximate
equalities'' subject to statistical fluctuations in the corresponding
empirical quantities, and an appropriate choice of the tolerances used
to determine equality, as functions of the input length $n$, is crucial
to turn the conceptual scheme into an effective algorithm. Specific
choices for tolerances are not discussed in~\cite{colt04}. The
attractive feature of the approach in~\cite{colt04} is its low
complexity; equipped with a reasonable choice of tolerance thresholds,
an efficient algorithm for the special case of processes of order one
can be implemented. However, as we shall see in the sequel, the
convergence of the algorithm is rather slow in practice, and very long
samples are necessary to achieve good deinterleaving performance,
compared to the schemes proposed here. The problem of deinterleaving
hidden-Markov processes was also studied, mostly experimentally,
in~\cite{Landwehr08}. Another variant of the problem, where all the
component processes are assumed to be identical (over the same
alphabet), of order one, and interleaved by a memoryless switch, was
studied in~\cite{Gillblad_et_al'2009}.

We note that IMPs are a special case of the broader class of
\emph{switching discrete sources} studied in~\cite{Shtarkov'92}, with
variants dating back as early as~\cite{Dobrushin'63}. However, the
emphasis in~\cite{Shtarkov'92} is on universally compressing the output
of a switched source of known structure, and not on the problem studied
here, which is precisely to identify the source's structure.

The rest of the paper is organized as follows. In
Section~\ref{sec:preliminaries} we present some additional definitions
and notation, and give a more formal and detailed definition of an IMP,
which will be useful in the subsequent derivations. We also show that
an IMP can be represented as a unifilar \emph{finite-state machine}
(FSM) source (see, e.g.,~\cite{Ash}), whose parameters satisfy certain
constraints induced by the IMP structure. In
Section~\ref{sec:uniqueness} we study conditions for uniqueness of the
IMP representation of a process. We identify two phenomena that may
lead to ambiguities: a so-called
\emph{alphabet domination} phenomenon which may arise from certain
transition probabilities in the switch being set to zero (and which,
therefore, does not arise in the case of memoryless switches), and the
presence of memoryless component processes. We derive a set of
sufficient conditions for uniqueness, and, in cases where ambiguities
are due solely to memoryless components (the so-called
\emph{domination-free} case, which includes all cases with memoryless
switches), characterize all the IMP representations of a process $P$.
 Most of the derivations and
proofs for the results of Section~\ref{sec:uniqueness} are presented in
Appendix~\ref{app:uniqueness}. In Section~\ref{sec:deinterleaving} we
present our deinterleaving scheme, establish its strong consistency,
and show that in the domination-free case, it can identify all valid
IMP representations of the interleaved process. The derivations and
proofs for these results are presented in
Appendix~\ref{app:consistency}. Finally, in
Section~\ref{sec:experimental} we present some experimental results for
practical implementations of deinterleaving schemes. We compare the
performance of our scheme with that of an implementation of the scheme
of~\cite{colt04} (with optimized tolerances) for the case of IMPs with
memoryless switches, showing that the ML-based deinterleaver achieves
high accuracy rates in identifying the correct alphabet partition for
much shorter sequences than those required by the scheme
of~\cite{colt04}. Our ideal scheme calls for finding the optimal
partition through an exhaustive search, which is computationally
expensive. Consequently, we show results for a randomized gradient
descent heuristic that searches for the same optimal partition.
Although in principle this approach sacrifices the optimality
guarantees of the ideal scheme, in practice, we obtain the same results
as with exhaustive search, but with a much faster and practical scheme.
We also present results for IMPs with switches of order one. We show,
again, that the ML-based schemes exhibit high deinterleaving success
rates for sequences as short as a few hundred symbols long, and perfect
deinterleaving, for the samples tested, for sequences a few thousand
symbols long.

\section{Preliminaries}\label{sec:preliminaries}

\subsection{Definitions}
All  Markov processes are assumed to be time-homogeneous and ergodic,
and, consequently, to define limiting stationary
distributions~\cite{Feller1}.
 We denote the (minimal) order of $P_i$ by $k_i\defined\ord(P_i)$,
refer to reachable strings $u^{k_i}$ as
\emph{states} of $P_i$, and denote the set of such states by
$\states(P_i)$, $i\in\onemw$. Some conditional probabilities may be
zero, and some $k_i$-tuples may be non-reachable, but all states are
assumed to be reachable and recurrent. We further assume that all
symbols $a\in\alphabet$ (and subalphabets $A\in\partn$) occur
infinitely often, and their stationary marginal probabilities are
positive. We make no assumptions on the initial conditions of each
process, and, in our characterization of ambiguities, distinguish
processes only up to their stationary distributions, i.e., we write
$P=P'$ if and only if $P$ and $P'$ admit the same stationary
distribution. All probability expressions related to stochastic
processes will be interpreted as (sometimes marginal) stationary
probabilities, e.g., $P_i(u)$, or $P_i(a|u)= P_i(ua)/P_i(u)$ when $u$
is not long enough to define a state of $P_i$. Aside from simplifying
some notations, this assumption makes our results on uniqueness of IMP
representations slightly stronger than if we had adopted a stricter
notion of process equivalence (e.g., actual process identity).

For a string $u^t=u_1 u_2 \ldots u_t \in \alphabet^t$, let
$\alphseq(u^t)\in\partn^t$ denote the corresponding string of
subalphabets, i.e., $\alphseq(u^t)_j = A_i$ where $i$ is the unique
index such that $u_j\in A_i \in\partn$, $1\le j \le t$. We sometimes
refer to $\alphseq(u^t)$ also as the \emph{switch sequence}
corresponding to $u^t$. Also, for $\alphabet'\subseteq\alphabet$, and a
string
    $u$ over $\alphabet$, let $\proj{u}{\alphabet'}$ denote the
    string over $\alphabet'$ obtained by deleting from $u$ all
    symbols that are not in $\alphabet'$.
The IMP $P=\IPi(P_1,P_2,\ldots,P_m;\pswitch)$ is formally defined as
follows:   Given $z^{t}\in\alphabet^{t}$, $t\ge 1$, and assuming
$z_t\in A_i$, we have
\begin{equation}\label{eq:IMP}
P(z_{t}|z^{t-1}) = \pswitch(A_i|\alphseq(z^{t-1}))P_i(z_{t}|z^{t-1}[A_i])\,.
\end{equation}
 It is readily verified
that~(\ref{eq:IMP}) completely defines the process $P$, which inherits
whatever initial conditions hold for the component and switch
processes, so that~(\ref{eq:IMP}) holds for any conditioning string
$z^{t-1}$, $t\ge 1$ (including $z^{t-1}=\lambda$).  Also, by recursive
application of~(\ref{eq:IMP}), after rearranging factors, we obtain,
for any sequence $z^n\in\alphabet^n$,
\begin{equation}\label{eq:P(z^n)}
P(z^n) = \pswitch(\alphseq(z^n))\prod_{i=1}^m P_i(z^n[A_i])\,.
\end{equation}
Notice that when initial conditions are such that the probabilities on
the right-hand side of~(\ref{eq:P(z^n)}) are stationary, the equation
defines a stationary distribution for $P$. (We adopt the convention
that $P_i(\lambda)=1$, $i\in\onemw$, and, consequently,
$P(\lambda)=1$.)

For conciseness, in the sequel, we will sometimes omit the arguments
from the notations $\IPi$ or $\IPip$, assuming that the respective sets
of associated subalphabets and processes (resp.\ $\{A_i\},\,\{P_i\}$ or
$\{A'_i\},\,\{P'_i\}$) are clear from the context. For IMP
representations $\IPi$ and $\IPip$, we write $\IPi\equiv\IPip$ if the
representations are identical, i.e., $\partn=\partnp$ and $P_i=P'_i$,
$i\in\onemw$ (in contrast with the relation $\IPi=\IPip$, which is
interpreted to mean that $\IPi$ and $\IPip$ generate the same process).

We will generally denote sequences (or strings) over $\alphabet$ with
lower case letters, e.g., $u\in\alphabet^\ast$, and sequences over
$\partn$ with upper case letters, e.g., $U\in\partn^\ast$. We say that
$u^n\in\alphabet^n$ and $U^n\in\partn^n$ are \emph{consistent} if
$P(u^n)>0$ and $U^n =\alphseq(u^n)$. Clearly, for every sequence $u^n$
with $P(u^n)>0$ there exists a sequence $U^n = \alphseq(u^n)$,  with
$\pswitch(U^n)>0$, that is consistent with $u^n$; conversely, if
$\pswitch(U^n)>0$, it is straightforward to construct sequences $u^n$
consistent with $U^n$. Unless specified otherwise, we assume that an
upper case-denoted alphabet sequence is consistent with the
corresponding lower case-denoted string, e.g., when we write
$UV=\alphseq(uv)$, we also imply that $U=\alphseq(u)$ and
$V=\alphseq(v)$.

\subsection{IMPs and FSM sources}
\label{sec:FSM}
A \emph{finite state machine} (FSM)  over an alphabet $\alpF$ is
defined by a triplet $F=(S,s_0,f)$, where $S$ is a set of
\emph{states}, $s_0\in S$ is a (possibly random)
\emph{initial state}, and $f:S\times\alpF\to S$ is a
\emph{next-state function}. A (unifilar) \emph{FSM source} (FSMS) is defined
by associating a conditional probability distribution $\PF(\cdot|s)$
with each state $s$ of $F$, and a probability distribution
$\PF^{\text{init}}(\cdot)$ on the initial state $s_0$. To generate a
random sequence $x^n$, the source draws $s_0$ according to
$\PF^{\text{init}}(\cdot)$ and then draws, for each $i$, $1\le i\le n$,
a symbol $x_i\in\alpF$ distributed according to $\PF(\cdot|s_{i-1})$,
and transitions to the state $s_i = f(s_{i-1},x_i)$. Markov sources of
order $k$ over $\alpF$ are special cases of FSMSs with $S =\alpF^k$. We
next observe that an \ISMP\ can be represented as an FSMS. For
convenience, we will assume in the discussion that FSMSs have arbitrary
but
\emph{fixed} initial states. In particular, we will assume that a fixed
initial state $s_0^{(j)}\in\states(P_j)$ is defined for the
component/switch processes $P_j$, $j\in\onemw$, where we recall that
$\states(P_j)$ denotes the state set of $P_j$. The results
 are easily generalized to arbitrary initial state conditions,
since any initial state distribution can be written as a convex
combination of fixed initial state conditions.

 We refer to the vector
$\bldk=(k_1,k_2,\ldots,k_m,\kswitch)$, where $k_j=\ord(P_j)$,
$j\in\onemw$,  as the
\emph{order vector} of the IMP $\IPi$. We denote by
$f_j$ the next-state function of the FSM associated with $P_j$,
$j\in\onemw$, and define the initial state vector $\blds_0 =
(s_0^{(1)},s_0^{(2)},\ldots,s_0^{(m)},s_0^{(\swltr)})$.  We consider
now an FSM $\FP = (S,
\blds_0,f)$, with state set $S = S_1\times S_2 \times \cdots S_m \times
S_{\swltr}$,  and next-state function $f$  defined as follows: Given a
state $\blds = (\sti{1},\sti{2},\ldots,\sti{m},\sti{\swltr})
\in S$, and $a\in
\alphabet$ such that $\alphseq(a)=A_i$, we have $f(\blds,a) = \blds' =
(\stip{1},\stip{2},\ldots,\stip{m},\stip{\swltr})$ where
$\stip{j}=\sti{j}$ for $j\in \onem\setminus \{i\}$,  $\stip{i} =
f_i(\sti{i},a)$, and $\stip{\swltr} = \fswitch(\sti{\swltr},A_i)$.
 To complete the definition of an FSMS,
for each state $\blds\in S$, we define the conditional probability
distribution
\begin{equation}\label{eq:PF}
\PFP(a \,|\,\blds) = \pswitch(A_i|\sti{\swltr})P_i(a\,|\,\sti{i}),
\;\;a\in \alp,\;\;\alphseq(a)=A_i\in\partn \,.
\end{equation}
The following proposition is readily verified.
\begin{proposition}\label{prop:FSM}
$\FP$, with transition probabilities
$\PFP$,~generates~$P{=}\IPi(P_1,P_2,\ldots,P_m,\pswitch)$.
\end{proposition}

Results analogous to Proposition~\ref{prop:FSM} for switching discrete
sources are given in~\cite{Shtarkov'92}. The class of finite state
sources considered in~\cite{Shtarkov'92}, however, is broader, as
unifilarity is not assumed.

It follows from the ergodicity and independence assumptions for IMP
components and switch that $P$ is an ergodic FSMS, and every state
$\blds\in S$ has a positive stationary probability. Let
$\alpha_i=|\alp_i|$, $1\leq i \leq m$, and
$\alpha=|\alphabet|=\sum_{i=1}^m\alpha_i$. By the definition of the
state set $S$, we have $|S|\le m^{k_\swltr}\prod_{i=1}^m
\alpha_i^{k_i}$ (equality holding when all $k_j$-tuples over the appropriate alphabet
are reachable states of $P_j$, $j\in\onemw$). Hence, the class of
\emph{arbitrary} FSMSs over $\alphabet$, with underlying FSM $\FP$, would have,
in general, up to
\begin{equation}\label{eq:FSMparams}
\dpartn(\partn,\bldk)
= (\alpha-1)m^{k_\swltr}\prod_{i=1}^m
\alpha_i^{k_i}
\end{equation}
 free statistical parameters. The conditional
probability distributions in~(\ref{eq:PF}), however, are highly
constrained, as the parameters $\PFP(a|\blds)$ satisfy relations of the
form
\begin{equation*}\label{eq:relations}
\pswitch(A_i|\stip{\swltr})\PFP(a|\blds) = \pswitch(A_i|\sti{\swltr})\PFP(a|\blds'),
\end{equation*}
where $A_i=\alphseq(a)$, for all states $\blds'$ such that
$\sti{i}=\stip{i}$. In particular, it follows directly
from~(\ref{eq:PF}) that $\PFP(a|\blds) =
\PFP(a|\blds')$ if $\sti{i}=\stip{i}$ and $\sti{\swltr}=\stip{\swltr}$.
Overall, the number of free parameters remains, of course, that of the
original component Markov processes and switch, i.e., up to
\begin{equation}\label{eq:IMPparams}
\dpartni(\partn,\bldk) =
\sum_{i=1}^m\alpha_i^{k_i}(\alpha_i-1) + (m-1)m^{k_\swltr},
\end{equation}
which is generally (much) smaller than $\dpartn(\partn,\bldk)$.

We refer to an FSMS satisfying the constraints implicit
in~(\ref{eq:PF}) as an
\emph{IMP-constrained} FSMS. Notice
that, given a specific IMP $P=\IPi(P_1,P_2,\ldots,P_m;\pswitch)$, the
associated FSM $\FP$ may also incorporate ``hard constraints'' on the
parameters of (maybe other) FSMSs based on $\FP$, due to some
$k_j$-tuples possibly being non-reachable in $P_j$, and the
corresponding transition probabilities being identically zero. Later
on, when our task is to estimate the FSMS without any prior knowledge
on the structure of $P$, we will assume that candidate structures $\FP$
are fully parametrized, i.e., the class of IMP-constrained FSMS
generated by $\FP$ has exactly $\dpartni$ free statistical parameters
(we omit the arguments of $\dpartn$ and $\dpartni$ when clear from the
context).

\section{Uniqueness of IMP representations}
\label{sec:uniqueness}

In this section, we study conditions under which the IMP representation
of a process is unique, and, for IMPs that are free from certain
``pathologies'' that will be discussed in the sequel, characterize all
IMP representations of a process when multiple ones exist. Notice that
although, as shown in Section~\ref{sec:preliminaries}, IMPs can be
represented as constrained FSM sources, the study of ambiguities of IMP
representations differs from the problem of characterizing different
FSM representations of a source~\cite{WF94}, or more generally of
representations of hidden Markov
processes~\cite{Blackwell-Koopmans'57}. It is known~\cite{WF94} that
all FSMs that
can generate a given FSMS $P$ are \emph{refinements}%
\footnote{
A refinement~\cite{FedMerGut'92} of an FSM $F=(S,s_0,f)$ is an FSM
$\refF=(S^\cref,s^\cref_0,f^\cref)$ such that for some fixed function
$g:S^\cref\to S$ and any sequence $x^n$, the respective state sequences
$\{s_i\}$ and $\{s^\cref_i\}$ satisfy $s_i = g(s^\cref_i)$, $0\leq i
\leq n$ (for example, the FSM underlying a Markov process of order $k+1$
is a refinement of the FSM underlying one of order $k$). By suitable
choices of conditional probabilities, a refinement of $F$ can generate
any process that $F$ can generate. } %
of a so called \emph{minimal FSM representation} of the source. In
particular, this applies to the FSM corresponding to any IMP
representation. However, the minimal FSM representation is not required
to satisfy the IMP constraints, so it needs not coincide with a minimal
(or unique) IMP representation. Notice also that, when defining IMPs
and their FSM representations, we have assumed that the orders $k_i$ of
all the Markov processes involved are minimal, thus excluding obvious
FSM refinements resulting from refining some of the individual Markov
processes.

\subsection{Alphabet domination}
\label{sec:domination}
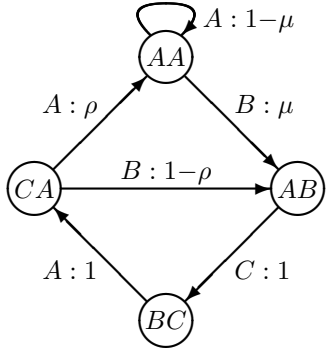
\begin{figure}
\centering{
\newcommand{\cdiam}{100}
\setlength{\unitlength}{0.070mm}
\begin{picture}(500,580)(0,0)
\thicklines
\put(0,250){\circle{\cdiam}}
\put(500,250){\circle{\cdiam}}
\put(250,0){\circle{\cdiam}}
\put(250,500){\circle{\cdiam}}
\put(0,250){\makebox(0,0){\small$CA$}}
\put(500,250){\makebox(0,0){\small$AB$}}
\put(250,0){\makebox(0,0){\small$BC$}}
\put(250,500){\makebox(0,0){\small$AA$}}
\put(220,542){
\put(0,0){\qbezier(0,0)(-60,60)(30,60)}
\put(0,0){\qbezier(30,60)(120,60)(60,0)}
\put(70,10){\vector(-1,-1){10}}
\put(0,0){\line(1,-1){20}}
}
\put(320,570){\makebox(0,0)[l]{\small$A:1{-}\mu$}}
\put(37,287){\vector(1,1){176}}
\put(120,405){\makebox(0,0)[r]{\small$A:\rho$}}
\put(287,463){\vector(1,-1){176}}
\put(380,405){\makebox(0,0)[l]{\small$B:\mu$}}
\put(463,213){\vector(-1,-1){176}}
\put(380,95){\makebox(0,0)[l]{\small$C:1$}}
\put(213,37){\vector(-1,1){176}}
\put(120,95){\makebox(0,0)[r]{\small$A:1$}}
\put(50,250){\vector(1,0){400}}
\put(250,280){\makebox(0,0){\small$B:1{-}\rho$}}
\end{picture}%
}
\caption{\label{fig:AB}A switch $\pswitch$ of order two over $\partn=\{A,B,C\}$.
Arcs are labeled $X:\xi$, where $X$ is the emitted symbol and $\xi$ the
corresponding transition probability.
Transitions not drawn are assumed to have probability zero.}
\vspace{-0.5cm}
\end{figure}

Let $A$, $B$ be arbitrary subalphabets in $\partn$. We say that $A$
\emph{dominates} $B$ (relative to $\pswitch$) if there exists a
positive integer $\LLL$ such that if $\pswitch$ has emitted $\LLL$
occurrences of $B$ without emitting one of $A$, then
\emph{with probability one} $\pswitch$ will emit an occurrence of $A$
before it emits another occurrence of $B$; in other words, if
$\pswitch(U)>0$, then $U[\{A,B\}]$ does not contain any run of more
than $\LLL$ consecutive occurrences of $B$. We denote the domination
relation of $A$ over $B$ as $A\dom B$, dependence on $\pswitch$ being
understood from the context; when $A$ does not dominate $B$, we write
$A\ndom B$ (thus, for example, $A\ndom A$). We say that $A$ is
\emph{\dominant} (in $\partn$, relative to $\pswitch$) if either $m=1$
(i.e., $\partn=\{A\}$) or $A\dom B$ for some $B\in\partn$, and that $A$
is \emph{\totdominant} if either $m=1$
 or $A\dom B$ for \emph{all} $B\in\partn\setminus\{A\}$.
 If $A\dom B$ and $B\dom A$, we say that $A$ and $B$ are in
 \emph{mutual domination}, and write $A \mdom B$.
It is readily verified that domination is an irreflexive transitive
relation. When no two subalphabets are in mutual domination, the
relation defines a strict partial order (see, e.g.,~\cite{Stanley1}) on
the finite set $\partn$. We shall make use of the properties of this
strict partial order in the sequel.

Domination can occur only if some transition probabilities in
$\pswitch$ are zero; therefore, it never occurs when $\pswitch$ is
memoryless. The approach for $\ord(\pswitch)=1$ in~\cite{colt04}
assumes that $\pswitch(A|A)>0$ for all $A\in\partn$. Clearly, this
precludes alphabet domination. However, the condition
 is too stringent to do so, or as a condition for uniqueness.

\begin{example}\label{ex:AB}  Consider an IMP $P=
\IPi(P_1,P_2,P_3;\pswitch)$ with $\partn=\{A,B,C\}$,
and $\pswitch$ as defined by Figure~\ref{fig:AB}, where
$\ord(\pswitch)=2$, and transitions are labeled with their respective
emitted symbols and probabilities. We assume that $\mu\in (0,1]$ and
$\rho\in (0,1)$.  For this switch, we have $A\dom B$, $A\dom C$, and
$B\mdom C$; $A$ is totally dominant, and, if $\mu < 1$, it is not
dominated. If $\mu=1$, every pair of subalphabets is in mutual
domination. In all cases, $\pswitch$ is aperiodic.
\end{example}

\subsection{Conditions for uniqueness}
\label{sec:conditions}
We derive sufficient conditions for the uniqueness of IMP
representations, and show how ambiguities may arise when the conditions
are not satisfied. The main result of the subsection is given in the
following theorem, whose derivation and proof are deferred to
Appendix~\ref{app:Th1&2}.

\begin{theorem}\label{th:ambiguities}
Assume that, for an IMP $P=\IPi(P_1,P_2,\ldots,P_m;\pswitch)$,
\begin{enumerate}
\renewcommand{\theenumi}{\roman{enumi}}
\item\label{as:noMD} no two subalphabets in
$\partn$ are in mutual domination,
\item\label{as:noTD} no subalphabet in $\partn$ is totally dominant, and
\item\label{as:noMM} none of the processes $P_i$ is memoryless.
\end{enumerate}
Then, if $P=\intlv_{\partn'}(P_1',P_2',\ldots,P_{m'};\pswitchp)$ for
some partition $\partnp$ and finite memory processes
$P_1',P_2',\ldots,P_{m'},\pswitchp$, we must have $\IPi\equiv\IPip$.
\iffalse
$\partn=\partnp$ and, thus, $m=m'$ and  $P_i=P_i'$, $i\in\onemw$.
\fi
\end{theorem}

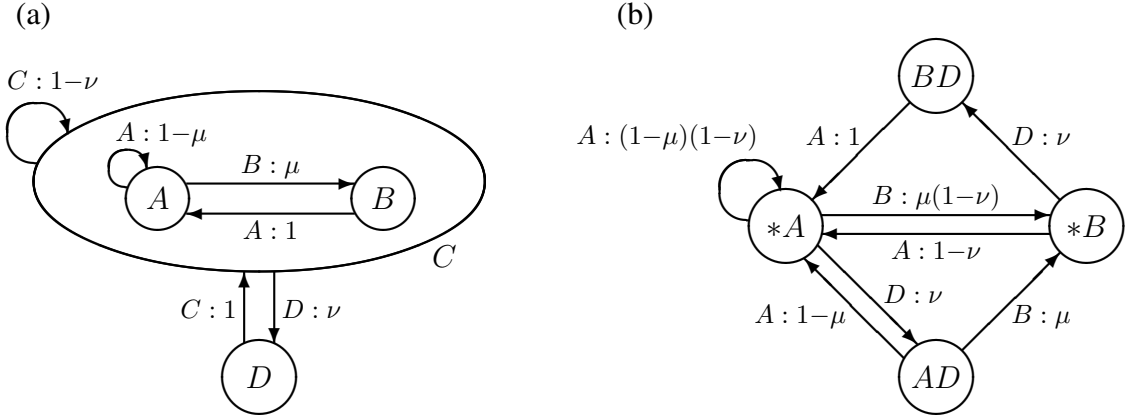
\begin{figure}
\centering{
{
\newcommand{\cdiam}{100}
\setlength{\unitlength}{0.1mm}
\begin{picture}(1450,550)(0,-50)
\thicklines
\put(0,480){\makebox(0,0){(a)}}
\put(0,80){
\put(0,60){
%\input{ellipse.tex}
% Ellipse: u = 300.0 v = 120.0 a = 300.0 b = 120.0 phi = 0.0 Grad
\qbezier(600.0, 120.0)(600.0, 169.7056)(512.132, 204.8528)
\qbezier(512.132, 204.8528)(424.2641, 240.0)(300.0, 240.0)
\qbezier(300.0, 240.0)(175.7359, 240.0)(87.868, 204.8528)
\qbezier(87.868, 204.8528)(0.0, 169.7056)(0.0, 120.0)
\qbezier(0.0, 120.0)(0.0, 70.2944)(87.868, 35.1472)
\qbezier(87.868, 35.1472)(175.7359, 0.0)(300.0, 0.0)
\qbezier(300.0, 0.0)(424.2641, 0.0)(512.132, 35.1472)
\qbezier(512.132, 35.1472)(600.0, 70.2944)(600.0, 120.0)
}
\put(-5,7){
\put(170,150){\circle{90}}
\put(170,150){\makebox(0,0){$A$}}
\put(130,190){\oval(50,50)[t]}
\put(130,190){\oval(50,50)[l]}
\put(151.75,209){\vector(1,-4){5}}
\put(175,235){\makebox(0,0){\small$A:1{-}\mu$}}
\put(470,150){\circle{90}}
\put(470,150){\makebox(0,0){$B$}}
\put(208,170){\vector(1,0){224}}
\put(320,193){\makebox(0,0){\small$B:\mu$}}
\put(432,130){\vector(-1,0){224}}
\put(320,108){\makebox(0,0){\small$A:1$}}
}
\put(5,244){
\put(0,0){\oval(80,80)[t]}
\put(0,0){\oval(80,80)[l]}
\put(37,18){\vector(1,-4){5}}
\put(-40,70){\makebox(0,0)[l]{\small$C:1{-}\nu$}}
}
\put(530,80){\makebox(0,0)[l]{$C$}}
\put(300,-80){\circle{100}}
\put(300,-80){\makebox(0,0){$D$}}
\put(280,-36){\vector(0,1){96}}
\put(270,10){\makebox(0,0)[r]{\small$C:1$}}
\put(320,60){\vector(0,-1){96}}
\put(330,10){\makebox(0,0)[l]{\small$D:\nu$}}
}
%
% Right side
%
\put(1000,0){
\put(-200,480){\makebox(0,0){(b)}}
\put(0,200){\circle{\cdiam}}
\put(-48,248){
\put(0,0){\oval(80,80)[t]}
\put(0,0){\oval(80,80)[l]}
\put(37,18){\vector(1,-4){5}}
\put(-230,70){\small\makebox(0,0)[l]{$A:(1{-}\mu)(1{-}\nu)$}}
}
\put(400,200){\circle{\cdiam}}
\put(200,0){\circle{\cdiam}}
\put(200,400){\circle{\cdiam}}
\put(0,200){\makebox(0,0){$\ast A$}}
\put(400,200){\makebox(0,0){$\ast B$}}
\put(200,0){\makebox(0,0){$AD$}}
\put(200,400){\makebox(0,0){$BD$}}
\put(43,175){\vector(1,-1){132}}
\put(157,25){\vector(-1,1){132}}
\put(47,215){\vector(1,0){305}}
\put(352,190){\vector(-1,0){305}}
\put(235,35){\vector(1,1){130}}
\put(365,235){\vector(-1,1){130}}
\put(165,365){\vector(-1,-1){130}}
\put(100,320){\makebox(0,0)[r]{\small$A:1$}}
\put(300,320){\makebox(0,0)[l]{\small$D:\nu$}}
\put(300,80){\makebox(0,0)[l]{\small$B:\mu$}}
\put(80,80){\makebox(0,0)[r]{\small$A:1{-}\mu$}}
\put(130,110){\makebox(0,0)[l]{\small$D:\nu$}}
\put(200,238){\makebox(0,0){\small$B:\mu(1{-}\nu)$}}
\put(200,167){\makebox(0,0){\small$A:1{-}\nu$}}
}
\end{picture}
} %
} %
\caption{\label{fig:ABCD}Switches for ambiguous IMP representation:
(a) $\pswitch$ over $\{C,D\}$, $\ord(\pswitch)=1$ ($C=A\cup B$, and the internal structure of $P_C$ is
also shown),
(b) $\pswitchp$ over $\{A,B,D\}$, $\ord(\pswitchp)=2$.
Arcs are labeled with their corresponding emitted symbols and transition probabilities;
transitions not shown have probability zero.}
\end{figure}

\begin{example}\label{ex:ambiguity}
We consider alphabets $A,B,D$, and $C=A\cup B$, and respective
associated processes $P_A,P_B,P_D,P_C$. Part (a) of Fig.~\ref{fig:ABCD}
shows a switch $\pswitch$ of order 1 over $\partn=\{C,D\}$. Here, $P_C$
is in itself an interleaved process $P_C=
\intlv_{\{A,B\}}(P_A,P_B;\pswitch^C)$ with
$P_B$ chosen as a memoryless process so that $P_C$ has finite memory
(specifically, $\ord(P_C)\le 2\,\ord(P_A)$); $P_D$ is not memoryless,
and we have $\nu,\mu\in(0,1)$. Part (b) shows a switch $\pswitchp$ of
order two over $\partnp=\{A,B,D\}$. State $\ast A$ (resp.\ $\ast B$)
represents all states that end in $A$ (resp.\ $B$). It is readily
verified that $P =
\IPi(P_C,P_D;\pswitch) =
\IPip(P_A,P_B,P_D;\pswitchp)$, so $P$ is an ambiguous IMP.
It is also readily verified that both $\IPi$ and $\IPip$ violate
Condition~(\ref{as:noTD}) of Theorem~\ref{th:ambiguities}: $C$ is
totally dominant in $\IPi$, and $A$ is totally dominant in $\IPip$. In
fact, the figure exemplifies a more detailed variant of
Theorem~\ref{th:ambiguities}, presented as Theorem~\ref{th:TD} below,
which characterizes ambiguities when Condition~(\ref{as:noTD}) of the
original theorem is removed.
\end{example}

Given partitions $\partn$ and $\partnp$ of $\alphabet$, we say that
$A_i\in\partn$ \emph{splits} in $\partnp$ if $A_i$ is partitioned into
subalphabets in $\partnp$, i.e. $A_j' \subseteq A_i$ for all
$A_j'\in\partnp$ such that  $A_j'
\cap A_i\nonempty$.

\begin{theorem}\label{th:TD}
Let $\partn = \{\alp_1,\alp_2,\ldots,\alp_m\}$ be a partition of
$\alphabet$, and consider an IMP representation
$P=\IPi(P_1,P_2,\ldots,P_{m};\pswitch)$ such that no two subalphabets
are in mutual domination, and none of the processes $P_i$ is
memoryless. Then, if $P=\IPip(P_1',P_2',\ldots,P_{m'}';\pswitchp)$ for
some partition $\partnp=\{A_1',A_2',\ldots,A_{m'}'\}$ of $\alphabet$,
we must have $A_i\in\partnp$ for all subalphabets $A_i\in\partn$ except
possibly for one subalphabet $A_{i_0}\in\partn$, which must be totally
dominant and split in $\partnp$.
\end{theorem}

The proof of Theorem~\ref{th:TD} is also deferred to
Appendix~\ref{app:Th1&2}. The theorem covers the special case $m=1$,
which is excluded by Condition~(\ref{as:noTD}) in
Theorem~\ref{th:ambiguities}. In this case, the IMP is actually a
finite-memory process, which admits the two ``obvious'' IMP
representations (with $m=1$ and $m=|\alphabet|=|A_1|$, respectively)
mentioned in the introduction.

\subsection{Ambiguities due to memoryless components in the domination-free case}
\label{sec:memoryless}
In this subsection, we eliminate Condition~(\ref{as:noMM}) of
Theorem~\ref{th:ambiguities},  while strengthening
Conditions~(\ref{as:noMD}) and~(\ref{as:noTD}) by excluding all forms
of alphabet domination. We characterize all the representations of an
IMP when ambiguities, if any, are due solely to memoryless components.

We say that a partition $\partnp$ is a
\emph{refinement} of $\partn$ if every subalphabet $A_i\in\partn$
splits in $\partnp$. When $\partnp$ is a refinement of $\partn$, we
define the function $\psimapfull:\partn'\to \partn $ mapping a
subalphabet $A_j'\in\partnp$ to the subalphabet $A_i\in\partn$ that
contains it. The notation and map extend in the natural way to
arbitrary strings, namely $\psimapfull:(\partnp)^k\to \partn^k$ for all
$k\ge 0$. We will omit the indices $\partn,\partnp$ from $\psimap$ when
clear from the context.
\begin{lemma}\label{lem:memoryless} Consider a partition
$\partn{\,=\,}\{\alp_1,\alp_2,\ldots,\alp_m\}$, together with a
refinement $\partnp=\{B_1,B_2,A_2,\ldots,A_m\}$ of $\partn$ (i.e.,
$A_1=B_1\cup B_2$). Let $P=\IPi(P_1,P_2,\ldots,P_m;\pswitch)$, where
$P_1$ is memoryless, and let
$P'=\IPip(\Ponej[1],\Ponej[2],P_2,\ldots,P_m;\pswitchp)$, where both
$\Ponej[1]$ and $\Ponej[2]$ are memoryless. Then, $P=P'$ if and only if
the following conditions hold:
\begin{equation}\label{eq:P'P''}
\Ponej[j](b)=\frac{P_1(b)}{P_1(B_j)},\;\;b\in B_j\,,\quad j\in\onetwo,
\end{equation}
\begin{equation}\label{eq:states'}
\states(\pswitchp)=\{\left.
S'\in(\partnp)^{\kswitch}\right|\psimap(S')\in\states(\pswitch)\},
\end{equation}
and for all $A\in\partnp$ and $S'\in\states(\pswitchp)$, with
$S=\psimap(S')$,
\begin{equation}\label{eq:splitswitch}
\pswitchp(A | S') = \begin{cases}
\pswitch(A | S),&A =A_i,\, i\ge 2,\\
\pswitch(A_1|S)P_1(B_j),& A = B_j, \, j=1,2\,.
\end{cases}
\end{equation}
\end{lemma}

\Remarks
The proof of Lemma~\ref{lem:memoryless} is deferred to
Appendix~\ref{app:memoryless}. The lemma is interpreted as follows:
since, given $\IPi$, processes $\Ponej[1],\Ponej[2]$, and $\pswitchp$
can always be defined to
satisfy~(\ref{eq:P'P''})--(\ref{eq:splitswitch}), an IMP $P$ with a
nontrivial memoryless component always admits alternative
representations where the alphabet associated with the memoryless
process has been split into disjoint parts (the split may be into more
than two parts, if the lemma is applied repeatedly). We refer to such
representations as
\emph{memoryless refinements} of the original representation
$\IPi$. Using the lemma repeatedly, we conclude that $P$ admits a
refinement where all the memoryless components are defined over
singleton alphabets. On the other hand, the memoryless components
$\Ponej[1]$ and $\Ponej[2]$ of $P'$ can be merged if and only if
$\pswitchp$ satisfies the constraint
\begin{equation}\label{eq:mergeconstraint}
 \pswitchp(B_2|S')=\gamma
\pswitchp(B_1|S')
\end{equation}
for a constant $\gamma$ independent of $S'\in\states(\pswitchp)$.
Indeed, when~(\ref{eq:mergeconstraint}) holds, we set $P_1(B_1) =
1/(1+\gamma)$ and $P_1(B_2)=\gamma/(1+\gamma)$, and $P_1,\pswitch$ are
defined implicitly by~(\ref{eq:P'P''})--(\ref{eq:splitswitch}). Notice
that the constraint~(\ref{eq:mergeconstraint}) is trivially satisfied
when the switch $\pswitchp$ is memoryless (and so is also the resulting
$\pswitch$). Thus, in this case, memoryless component processes can be
split or merged arbitrarily to produce alternative IMP representations.
When the switch has memory, splitting is always possible, but merging
is conditioned on~(\ref{eq:mergeconstraint}). We refer to a
representation where no more mergers of memoryless processes are
possible, as well as to the corresponding partition $\partn$,  as
 \emph{canonical} (clearly, the canonicity of $\partn$ is relative to the given IMP).\footnote{
 The particular case of this result for IMPs with memoryless
 switches discussed in~\cite{SSW-isit}
  uses a slightly different definition of
  canonicity.
} %

We denote the canonical representation associated with an IMP $P=\IPi$
by $(\IPi)^\ast$, and the corresponding canonical partition by
$(\partn)^\ast_P$. Also, we say $P$ is
\emph{\domfree}\ if there is no alphabet domination in \emph{any} IMP
representation of $P$. The main result of the subsection is given in
the theorem below, whose proof is presented in
Appendix~\ref{app:memoryless}.

\begin{theorem}\label{theo:memoryless}
Let $P=\IPi$ and $P'=\IPip$ be domination-free IMPs over $\alphabet$.
Then, $P=P'$ if and only if $(\IPi)^\ast\equiv(\IPip)^\ast$.
\end{theorem}

Theorem~\ref{theo:memoryless} implies that, in the domination-free
case, all the IMP representations of a process are those constructible
by sequences of the splits and mergers allowed by
Lemma~\ref{lem:memoryless}. In particular, this always applies to the
case of memoryless switches, where domination does not arise.
\begin{corollary}\label{cor:memoryless}
Let $P=\IPi$ and $P'=\IPip$ be IMPs over $\alphabet$, where the
switches $\pswitch$ and $\pswitchp$ are memoryless. Then, $P=P'$ if and
only if $(\IPi)^\ast\equiv(\IPip)^\ast$.
\end{corollary}

\section{The deinterleaving scheme}
\label{sec:deinterleaving}

Given any finite alphabet $A$, a sequence $u^t \in A^t$, and a
nonnegative integer $k$, denote by $\hhk(u^t)$ the $k$th order
(unnormalized) empirical entropy of $u^t$, namely, $\hhk(u^t) = -\log
\pmlk(u^t), $ where $\pmlk(u^t)$ is the ML (or empirical) probability
of $u^t$ under a $k$th order Markov model with a fixed initial state.
Let $z^n$ be a sequence over $\alphabet$. An arbitrary partition
$\partn$ of $\alphabet$ naturally defines a deinterleaving of $z^n$
into sub-sequences $\bldz_i = z^n[A_i],\quad 1\le i\le m, $ with a
switch sequence $\bldZ_{\swltr} =
\alphseq(z^n)$.
Given, additionally, an order vector $\bldk=(k_1,k_2,\ldots,k_m,k_\swltr)$,
we define
\begin{equation*}
\hhF(z^n) = {\sum_{i=1}^m }\hhk[k_i](\bldz_i)+ \hhk[k_\swltr](\bldZ_{\swltr})\,.
\end{equation*}
This quantity can be regarded as the (unnormalized) empirical entropy
of $z^n$ with respect to $F=\FP$ \emph{for an IMP-constrained} FSMS (as
discussed in Subsection~\ref{sec:FSM}). Indeed, let $\hpF(z^n)$ denote
the ML probability of $z^n$ with respect to $F$ under IMP constraints,
i.e., denoting by $\mcPFPI$ the class of all IMPs generated by $F$
(i.e., all FSMSs based on $F$ with parameter vectors satisfying the IMP
constraints), we have
\begin{equation}\label{eq:defineML}
\hpF(z^n) = \max_{P\in\mcPFPI} P(z^n)\,.
\end{equation}
Clearly, by~(\ref{eq:P(z^n)}), $\hpF(z^n)$ is obtained by maximizing,
independently, the probabilities of the component and switch sequences
derived from $z^n$, and, thus, we have $\hhF(z^n)=-\log\hpF(z^n)$.
Notice that $\hpF(z^n)$ is generally different from (and upper-bounded
by) the ML probability with respect to $F$ for an
\emph{unconstrained} FSMS; this ML probability will be denoted
$\pmlF(z^n)$. Next, we define the
\emph{penalized cost} of $z^n$ relative to $\partn$ and $\bldk$ as
 \begin{eqnarray}\label{eq:costpartn}
\cost_{\partn,\bldk}(z^n) &=& \hhF(z^n) + \beta \dpartni \log(n+1)\,,
\end{eqnarray}
where $\dpartni=\dpartni(\partn,\bldk)$, as given
in~(\ref{eq:IMPparams}), is the number of free statistical parameters
in a generic IMP-constrained FSMS based on $F$,
and $\beta$ is a nonnegative (penalization) constant.%
\footnote{
For convenience, we set the penalty terms in~(\ref{eq:costpartn}) all
proportional to $\log(n+1)$, rather than the term corresponding to
$\bldz_i$ being proportional to $\log |\bldz_i|$. Given our basic
assumptions on switch processes, if $z^n$ is a sample from an IMP,
$|\bldz_i|$ will, almost surely, be proportional to $n$. Therefore, the
simpler definition adopted has no effect on the main asymptotic
results. Clearly, using $\log(n+1)$ in lieu of $\log n$, which will be
convenient in some derivations, is also of negligible effect.}

 Given a sample $z^n$ from an IMP $P$, our deinterleaving scheme estimates a partition
 $\hat{\partn}(z^n)$, and an order vector $\hat{\bldk}(z^n)$,
for the estimated IMP representation of $P$.
 The desired estimates are obtained by the following rule:
\begin{equation}\label{eq:rulePi}
\left(\hat{\partn}(z^n),\hat{\bldk}(z^n)\right) = \arg\min_{(\partnp,\bldk')}\kcost[\partnp,\bldk'](z^n),
\end{equation}
where $(\partnp,\bldk')$ ranges over all pairs of partitions of
$\alphabet$ and order vectors $\bldk'$. In the minimization, if
$\kcost[\partnp,\bldk'](z^n)=\kcost[\partnpp,\bldk''](z^n)$, for
different pairs $(\partnp,\bldk')$ and $(\partnpp,\bldk'')$, the tie is
broken first in favor of the partition with the smallest number of
alphabets. Notice that although the search space in~(\ref{eq:rulePi})
is defined as a Cartesian product, once a partition $\partnp$ is
chosen, the optimal process orders $k'_j$ are determined independently
for each $j\in\onemw$, in a conventional penalized ML Markov order
estimation procedure (see, e.g.,~\cite{CsiszarShields00}). Also, it is
easy to verify that the optimal orders $\hat{k}_j$ must be $O(\log n)$,
reducing the search space for $\bldk'$ in~(\ref{eq:rulePi}).

Our main result is given by the following theorem, whose derivation and
proof are presented in Appendix~\ref{app:consistency}. Recall that
$\canon[P]{\partn}$ denotes the canonical partition of $P$
(Subsection~\ref{sec:memoryless}).
\begin{theorem}\label{th:consistency}
Let $P=\IPi(P_1,P_2,\ldots,P_m;\pswitch)$, and let $z^n$ be a sample
from $P$. Then, for suitable choices of the penalization constant
$\beta$, $\hat{\partn}(z^n)$ is compatible with $P$, and
$\hat{\bldk}(z^n)$ reproduces the order vector of the corresponding IMP
representation $\intlv_{\hat{\partn}}$, almost surely as $n\to\infty$.
Furthermore, if $P$ is
\domfree, we have
\[
\hat{\partn}(z^n) = \canon[P]{\partn}\quad\text{a.s.\ \   as}\; n\to\infty\,.
\]
\end{theorem}

\Remarks
\begin{itemize}
\item Theorem~\ref{th:consistency} states that our scheme, when presented
with a sample from an interleaved process, will almost surely recover
an alphabet partition compatible with the process. If the interleaved
process is \domfree, the scheme will recover the canonical partition of
the process, from which \emph{all} compatible partitions can be
generated via repeated applications of Lemma~\ref{lem:memoryless}. The
difficulty in establishing the first claim of the theorem resides in
the size of the class of models that participate in the
optimization~(\ref{eq:rulePi}). The fact that a compatible partition
will prevail over any specific incompatible one eventually almost
surely, for any penalization coefficient $\beta\ge 0$, will be readily
established through a large deviations argument. However, the class
contains models whose size is not bounded with $n$. In fact, it is well
known (see, e.g.,~\cite{cover}) that the stationary distribution of the
ergodic process $P$ can be approximated arbitrarily (in the entropy
sense) by finite memory processes of unbounded order. Thus, without
appropriately penalizing the model size, a sequence of ``single
stream'' hypotheses of unbounded order can get arbitrarily close in
cost to the partitions compatible with $P$. We will prove that an
appropriate positive value of $\beta$ suffices to rule out these large
models that asymptotically approach $P$. To establish the second claim
of the theorem, we will take advantage of the observation that the
canonical representation of a domination-free IMP, is, in a sense, also
the most ``economical''. Indeed, comparing the number of free
statistical parameters in the two IMP representations considered in
Lemma~\ref{lem:memoryless}, we obtain, using~(\ref{eq:IMPparams}),
\begin{equation}\label{eq:comparekappa}
\dpartni(\partnp,\bldk')-\dpartni(\partn,\bldk) = m(m+1)^{\kswitch}-(m-1)m^{\kswitch}-1\,.
\end{equation}
It is readily verified that the expression on the right hand side
of~(\ref{eq:comparekappa}) vanishes for $\kswitch=0$, and is strictly
positive when $\kswitch>0$ (since $m\ge 1$). Therefore, splitting a
memoryless component as allowed by Lemma~\ref{lem:memoryless}, in
general, can only increase the number of parameters. Thus, the
canonical partition minimizes the model size, and with an appropriate
choice of $\beta>0$, our penalized ML scheme will correctly identify
this minimal model.

\item If a bound is known on the orders of the component and
switch processes, then it will follow from the proof in
Appendix~\ref{app:consistency} that the first claim of
Theorem~\ref{th:consistency} can be established with any $\beta\ge
0$. However, an appropriate positive value of $\beta$ is still
needed, even in this case, to recover the canonical partition in
the second claim of the theorem. As mentioned, our deinterleaving
scheme assumes that IMPs based on $\FP$ are fully parametrized,
i.e., the class has $\dpartni$ free statistical parameters. If the
actual IMP being estimated is less than fully parametrized (i.e.,
it does have some transition probabilities set to zero), the effect
of penalizing with the full $\dpartni$ is equivalent to that of
using a larger penalization coefficient~$\beta$.
\end{itemize}

\section{Experimental results}\label{sec:experimental}

We report on experiments showing the performance of practical
implementations of the proposed deinterleaver. The experiments were
based on test sets consisting of $200$ interleaved sequences each. Each
sequence was generated by an IMP with $m{=}3$, subalphabet sizes
$\alpha_1{=}4,\,\alpha_2{=}5,\,\alpha_3{=}6$, component Markov
processes of order $k_i\le 1$  with randomly chosen parameters, and a
switch of order $\kswitch \le 1$ as described below. In all cases, the
switches were domination-free. Deinterleaving experiments were run on
prefixes of various lengths of each sequence, and, for each prefix
length, the fraction of sequences correctly deinterleaved was recorded.

In the first set of experiments, the component Markov processes, all of
order one, were interleaved by uniformly distributed memoryless
switches (i.e., $\kvec=(1,1,1,0)$). We compared the deinterleaving
performance of the ML-based scheme proposed here with that of an
implementation of the scheme of~\cite{colt04}, with tolerances for the
latter optimized (with knowledge of the correct partition) to obtain
the best performance for each sequence length. Two variants of the
ML-based scheme were tested: Variant (a) implements~(\ref{eq:rulePi})
via exhaustive search over all
partitions.\footnote{%
We recall that given a sequence $z^n$ and a partition $\partn$, the
order vector $\bldk$ minimizing the cost $\cost_{\partn,\bldk}(z^n)$ is
determined through conventional penalized-ML order estimators for the
various sub-sequences induced by $\partn$. We assume that this
minimizing order vector is used in all cost computations, and omit
further mention of it.}
 Since this is rather slow, a heuristic Variant (b) was developed, based on
 a randomized gradient descent-like search. This variant, which is briefly
 described next, is much faster, and
achieves virtually the same deinterleaving performance as the full
search.

\begin{table}
\caption{\label{tab:results}Fraction of correctly deinterleaved
sequences (out of $200$) vs.\ sequence length, for two variants of the
proposed scheme (ML({\rm a}) and ML({\rm b})), and for the scheme
of~\cite{colt04}. A penalization constant $\beta=\half$ was used in all
cases for the ML-based schemes.}
\begin{center}
\begin{tabular}{|r|rrr|rr|rr|}
\hline
    &\multicolumn{3}{|c|}{memoryless switch}&\multicolumn{4}{|c|}{switch with memory} \\
\cline{5-8}
    &\multicolumn{3}{|c|}{$\kvec=(1,1,1,0)$}&\multicolumn{2}{|c|}{$\kvec=(1,1,1,1)$} & \multicolumn{2}{|c|}{$\kvec=(0,1,1,1)$} \\
\cline{2-8}
   $$  &         &        &               &        &        & ML (b)$\;$ & ML (b)$\;$\\[-0.75em]
   $n$  & ML (a) & ML (b) & \cite{colt04} & ML (a) & ML (b) & canonical & compatible \\
\hline
250  & 0.010 & 0.010 & 0.000 & 0.310 & 0.300 & 0.215 & 0.225\\
500  & 0.135 & 0.130 & 0.000 & 0.635 & 0.620 & 0.600 & 0.625 \\
1000 & 0.440 & 0.420 & 0.000 & 0.915 & 0.915 & 0.880 & 0.900\\
2500  & 0.820 & 0.815 & 0.000 & 0.995 & 0.995 & 0.990& 0.990 \\
5000  & 0.960 & 0.960 & 0.005 & 1.000 & 1.000 & 1.000 & 1.000 \\
10000  & 0.990 & 0.990 & 0.030 & 1.000 & 1.000& 1.000 & 1.000 \\
15000  & 1.000 & 1.000 & 0.080 & 1.000 & 1.000& 1.000 & 1.000 \\
20000  & 1.000 & 1.000 & 0.135 & 1.000 & 1.000& 1.000 & 1.000 \\
50000  &       & 1.000 & 0.460 &       & 1.000& 1.000 & 1.000 \\
100000 &       & 1.000 & 0.770 &       & 1.000& 1.000 & 1.000 \\
500000 &       & 1.000 & 0.965 &       & 1.000& 1.000 & 1.000 \\
1000000&       & 1.000 & 0.980 &       & 1.000& 1.000 & 1.000 \\
\hline
\end{tabular}
\end{center}
\end{table}

We define the
\emph{neighborhood} of radius $t$ of a partition $\partn$, denoted
$\neighbor_t(\partn)$, which consists of all partitions $\partnp$
obtained from $\partn$ by switching up to $t$ symbols of $\alphabet$
from their original subalphabets in $\partn$ to other subalphabets
(including possibly new subalphabets not present in $\partn$). The main
component of the heuristic starts from an input sequence $z^n$ and a
random partition $\partn_0$ of $\alphabet$, and exhaustively searches
for the partition $\partnp$ that minimizes the cost
$\cost_{\partnp}(z^n)$ within the neighborhood $\neighbor_t(\partn_0)$,
for some small fixed value of $t$. The minimizing partition then
becomes the center for a new exhaustive neighborhood search. This
``greedy'' deterministic process continues until no improvements in the
cost function can be obtained. At this point, the best partition
$\partn$ observed so far is perturbed by picking a random partition
$\partnp_0\in\neighbor_r(\partn)$, for a fixed radius $r
> t$, and the deterministic search is repeated using $\partnp_0$ in lieu of $\partn_0$ as
the starting point. The routine stops if a given number $N$ of
consecutive rounds of such perturbations do not yield further cost
reductions, at which point the best partition $\partn$ observed so far
is returned as a candidate solution. To improve deinterleaving
reliability, this basic scheme can be run for several independent
starting random partitions $\partn_0$, noting the overall cost minimum.
The number $R$ of such outer iterations, the maximum count $N$ of
consecutive perturbations without improvement, and the neighborhood
radii $t$ and $r$, are parameters controlling the complexity vs.\
deinterleaving performance trade-off of the heuristic. For our
experiments, we found that $R=5$, $N=15$, $t=1$, and $r=2$, yielded
performance virtually identical to a full exhaustive partition search,
with orders of
magnitude reduction in complexity.\footnote{%
In fact, to keep running times reasonable, the exhaustive search was
given the benefit of limiting the search space to partitions $\partn$
with $|\partn| \le 4$. No such limitation was assumed for the heuristic
scheme, whose search space included, in principle, partitions of any
size $|\partn| \le |\alphabet|$.
} %

\begin{figure}
\begin{center}
\includegraphics[width=4in]{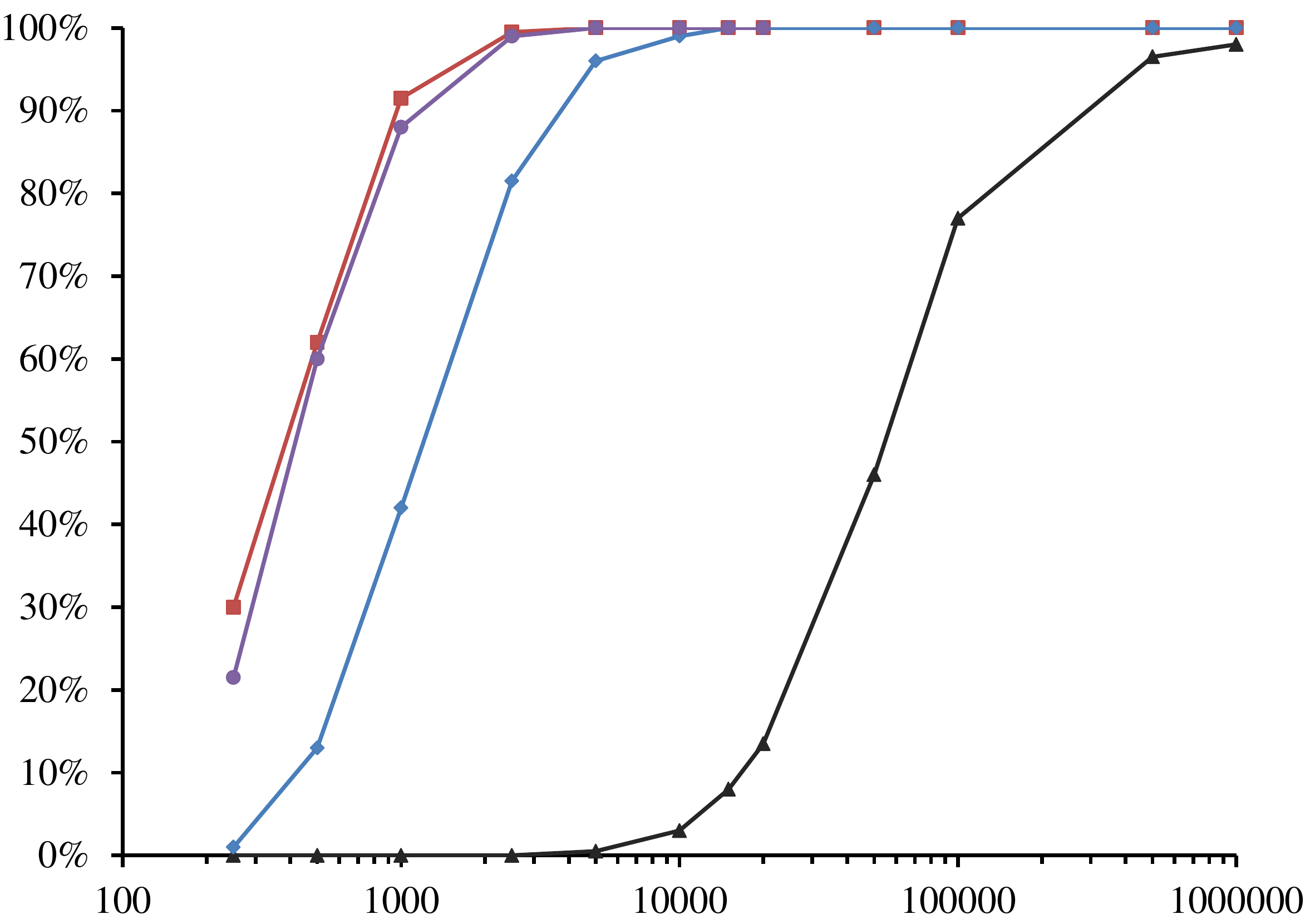}
\setlength{\unitlength}{0.01in}
\begin{picture}(0,0)(10,-20)
\put(0,0){$n$}
\put(-3,0){
\put(-56,220){\makebox(0,0){\textcircled{\footnotesize 2}}}
\put(-220,220){\makebox(0,0){\textcircled{\footnotesize 1}}}
\put(-260,220){\makebox(0,0){\textcircled{\footnotesize 4}}}
\put(-292,220){\makebox(0,0){\textcircled{\footnotesize 3}}}
}
\put(-105,110){%
\put(0,0){\put(0,0){\makebox(0,0){\textcircled{\footnotesize 1}}}%
\put(15,0){\makebox(0,0)[l]{\footnotesize $\kvec{=}(1,1,1,0)$ ML}}}%
\put(0,-20){\put(0,0){\makebox(0,0){\textcircled{\footnotesize 2}}}%
\put(15,0){\makebox(0,0)[l]{\footnotesize $\kvec{=}(1,1,1,0)$~\cite{colt04}}}}%
\put(0,-40){\put(0,0){\makebox(0,0){\textcircled{\footnotesize 3}}}
\put(15,0){\makebox(0,0)[l]{\footnotesize $\kvec{=}(1,1,1,1)$ ML}}}%
\put(0,-60){\put(0,0){\makebox(0,0){\textcircled{\footnotesize 4}}}
\put(15,0){\makebox(0,0)[l]{\footnotesize $\kvec{=}(0,1,1,1)$ ML}}}%
}
\put(-420,10){
\put(0,0){\begin{sideways}{\footnotesize Fraction of sequences correctly deinterleaved}\end{sideways}}
}
\end{picture}
\end{center}
\caption{\label{fig:plots}Deinterleaving success rate vs. sequence length for various IMPs and deinterleavers.}
\end{figure}
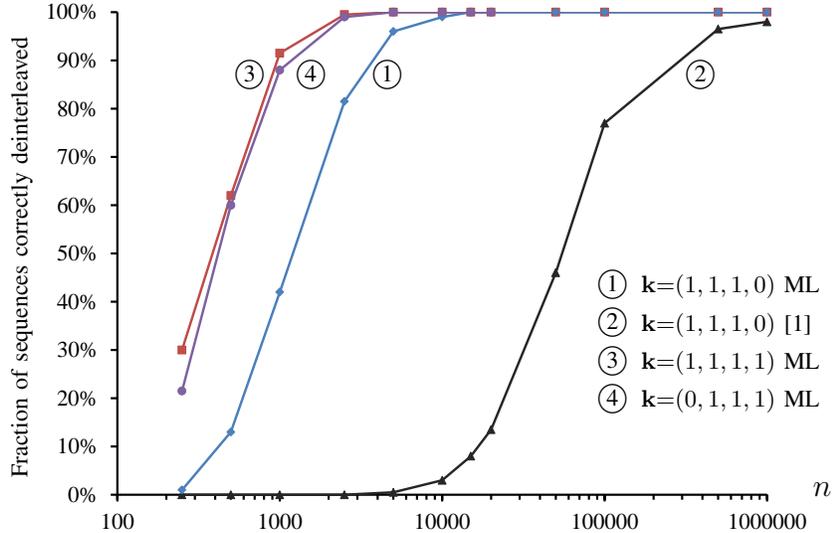

The results of the experiments with memoryless switches are summarized
in columns 2--4 of Table~\ref{tab:results}. The table shows that the
proposed ML-based scheme (in either variant) achieves better than
$80\%$ deinterleaving accuracy for sequences as short as $n=2500$, with
perfect deinterleaving for $n\ge 15000$, whereas the scheme
of~\cite{colt04}, although fast, requires much longer sequences,
correctly deinterleaving just one sequence in $200$ for $n=5000$, and
achieving $98\%$ accuracy for $n = 10^6$ (the maximum length tested in
the experiments). This comparison is illustrated by the curves labeled
\textcircled{\small 1} and \textcircled{\small 2} in Figure~\ref{fig:plots}.

In the second set of experiments, we used, for each sequence, the same
component processes as in the first set, but with a switch $\pswitch$
of order one (i.e., $\kvec=(1,1,1,1)$), with random parameters and
uniform marginal subalphabet probabilities. The results are presented
in columns 4--5 of Table~\ref{tab:results}, and plotted in the curve
labeled \textcircled{\small 3} in Figure~\ref{fig:plots}. We observe
that the additional structure resulting from the switch memory allows
for improved deinterleaving performance for shorter sequences: better
than $60\%$ accuracy is obtained for sequences as short as $n = 500$,
while perfect deinterleaving is obtained for $n \ge 5000$. A comparison
with the scheme of~\cite{colt04} is omitted in this case, as the
determination of appropriate statistics thresholds (not discussed
in~\cite{colt04}) appears more involved than in the memoryless switch
case, and is beyond the scope of this paper.

Finally, in a third set of experiments, we maintained switches of order
one, but let the component process $P_1$ in each case be memoryless
(i.e., $\kvec=(0,1,1,1)$). Recall that, by Lemma~\ref{lem:memoryless},
the resulting IMPs in this case have ambiguous representations. Results
for the heuristic ML-based scheme are presented in columns 6--7 of
Table~\ref{tab:results}, which list the fraction of sequences of each
length for which the deinterleaver picked the canonical partition, or
any compatible partition, respectively. We observe that, except for
minor deviations for the shorter sequence lengths, the deinterleaver
consistently picks the canonical partition, as expected from
Theorem~\ref{th:consistency}. The fraction of sequences for which the
canonical partition is chosen is plotted in the curve labeled
\textcircled{\small 4} in  Figure~\ref{fig:plots}. Memoryless
components are excluded in~\cite{colt04}, so a comparison is not
possible in this case.

Recalling the second remark at the end of
Section~\ref{sec:deinterleaving}, we note that any nonnegative value of
the penalization constant $\beta$ would have sufficed for the ML
schemes in the first two sets of experiments, since the IMPs considered
have unique representations, and the order of all the processes tested
was bounded by $1$. However, a positive value of $\beta$ is required to
recover the canonical partition (and from it, all compatible
partitions) in the case of the third set. For shorter sequences, a
value of $\beta$ as small as possible is preferred to exclude
non-compatible partitions, while a value of $\beta$ as large as
possible is preferred to recover the canonical partition. Overall, a
value $\beta=\half$ worked well in practice in all cases, providing the
best trade-off for shorter sequence lengths (clearly, the choice
becomes less critical as the sequence length increases). This value of
$\beta$ is smaller than the value employed in the proof of
Theorem~\ref{th:consistency}. In general, the question of determining
the minimal penalty that guarantees consistent deinterleaving remains
open. The situation bears some similarity to the one encountered with
Markov order estimators: while it is known that $\beta=\half$
guarantees strong consistency in all cases, it is also known that much
smaller penalization constants (or even penalization functions $o(\log
n)$) may suffice when the process order is
bounded~\cite{CsiszarShields00}. The general question of the minimal
penalization that guarantees consistent unbounded order estimation is,
also in this case, open~\cite{CsiszarShields00}.

\appendices

\section{Uniqueness of IMP representations: derivations}\label{app:uniqueness}
\subsection{Derivation of Theorems~\ref{th:ambiguities}
and~\ref{th:TD}}\label{app:Th1&2}

 Theorems~\ref{th:ambiguities}
and~\ref{th:TD} will be established through a series of lemmas.  The
first one (Lemma~\ref{lem:ergod} below) captures some essential
properties of the interleaved process
$P{=}\IPi(P_1,P_2,\ldots,P_m;\pswitch)$ and of the domination relation,
which we will draw upon repeatedly in the sequel. These properties
follow immediately from our ergodicity and independence assumptions.
Intuitively, the key point is that if $A_1\ndom A_2$, the interleaved
system can always take a trajectory (of positive probability) where it
reaches an arbitrary state $s$ of $P_1$, and then, without returning to
$A_1$, visits any desired part of $A_2$ any desired number of times
(while the state of $P_1$ remains, of course, unchanged). The last
segment of the trajectory, with an unbounded number of occurrences of
$A_2$, can be chosen independently of $s$. For ease of reference, these
observations are formally stated in the lemma below, where $\nocc{a}(z)
$ denotes the number of occurrences of a symbol $a$ in a string $z$.
\begin{lemma}
\label{lem:ergod} %
Consider the subalphabets $A_1,A_2\in\partn$, and assume $A_1\ndom
A_2$.
\renewcommand{\theenumi}{\roman{enumi}}
\begin{enumerate}
\item\label{it:UV}
Let $M_1$ and $\LLL$ be arbitrary integers. There exist strings
$U,V\in\partn^\ast$ such that $\pswitch(UV)>0$, $\nocc{A_1}(U) \ge
M_1$, $\nocc{A_1}(V) = 0$, $\nocc{A_2}(V) \ge
\LLL$, and $\pswitch(A_1\,|\,UV) > 0$.
\item\label{it:s}
Let $\MMM$ be an arbitrary integer, let $s$ be an arbitrary state
of $P_1$, and consider an arbitrary subset $B_2\subseteq A_2$ and
an integer $M_1 \ge k_1$. There exists an integer $\LLL \ge \MMM$,
and strings $u,v\in\alphabet^\ast$ such that $uv$ is consistent
with $UV$ (with $|u|=|U|$), where $U$ and $V$ are the strings
obtained from Part~\ref{it:UV}) for these values of $M_1$ and
$\LLL$, $u[A_1] = u's$ for some $u'\in A_1^\ast$,
$\bigstrlen{v[B_2]}\ge\MMM$, and the choice of $v$ does not depend
on $s$ (in particular, the same $v$ can be chosen for any
$s\in\states(P_1)\,$).
\end{enumerate}
\end{lemma}
\begin{proof}
Part~\ref{it:UV}) follows from the ergodicity of $\pswitch$, the
positivity of both $\pswitch(A_1)$ and $\pswitch(A_2)$, and the
definition of domination. The existence of the desired string $u$ in
Part~\ref{it:s}) follows further from the independence of the component
and switch processes, and from the ergodicity of $P_1$ (in particular,
the fact that $P_1(s)>0$). Relying also on the ergodicity of $P_2$, we
obtain the string $v$. The value of $M$ is determined by how many times
$v$ must visit $A_2$ to obtain $\MMM$ occurrences of symbols in the
subset $B_2$.  The independence of $v$ from $s$ follows
from~(\ref{eq:P(z^n)}), which allows us to substitute any string over
$A_1$, of positive probability, for $u[A_1]$ in $uv$, resulting in a
string $\tilde{u}v$, with $P(\tilde{u}v)>0$, $\tilde{u}$ compatible
with $U$, and $\tilde{u}[A_1]$ ending in any desired state of $P_1$.
\end{proof}

For succinctness, in the series of lemmas and corollaries that follows,
we assume \emph{throughout} that we are given an ambiguous IMP,
$P=\IPi(P_1,P_2,\ldots,P_{m};\pswitch)
=\intlv_{\partn'}(P'_1,P'_2,\ldots,P'_{m'};\pswitch')$, where
$\partn=\{\alp_1,\alp_2,\ldots,\alp_m\}$ and
$\partn'=\{\alp'_1,\alp'_2,\ldots,\alp'_{m'}\}$ are partitions of
$\alphabet$, with $\partn\ne\partnp$. Clearly, for at least one
alphabet $A_i$ we must have $A_i\not\in
\partnp$, so we assume, without loss of generality, that
$A_1\not\in\partnp$, and, furthermore, that $A_1\cap A_1' \nonempty$.
Also,  we say that two subalphabets $A_i,A_j\in\partn$
\emph{share} a subalphabet $A_{\ell}'\in\partnp$ if $A_{\ell}'$ intersects
both $A_i$ and $A_j$.

\begin{lemma}\label{lem:domination}
Assume that $A_2$ shares $A_1'$ with $A_1$, and $A_1\ndom A_2$. Then,
for all $a\in A_1\cap A_1'$, $P_1(a\,|\,s)$ is independent of
$s\in\states(P_1)$.
\end{lemma}
\ifproofs
\begin{proof}
Let $a\in A_1\cap A_1'$, and $s\in\states(P_1)$. Let
$U,V\in\partn^\ast$ and $u,v\in\alphabet^\ast$ be the strings
guaranteed by Lemma~\ref{lem:ergod} for the given state $s$, $\MMM =
\ord(P_1')$, and $B_2 = A_2
\cap A_1'$. Recall that $v$ can be chosen independently
of $s$, and $\bigstrlen{v[B_2]} \ge \MMM = \ord(P_1')$. Let
$\hat{v}=v[A_1']$, and let $U'V'=\alphseq[\partnp](uv)$. Then,
applying~(\ref{eq:IMP}) separately to each of the two given IMP
representations of $P$, and noting that $|\hat{v}|
\ge \bigstrlen{v[B_2]} \ge \ord(P_1')$, we have
\[
P(a|uv) = P_1(a|s)\pswitch(A_1|UV) = P_1'(a|\hat{v})\pswitchp(A_1'|U'V').
\]
Now, recalling that $\pswitch(A_1|UV)>0$ by
Lemma~\ref{lem:ergod}(\ref{it:UV}), we obtain
\[
P_1(a|s) = \frac{P_1'(a|\hat{v})\pswitchp(A_1'|U'V')}{\pswitch(A_1|UV)},
\]
which is independent of $s$.
\end{proof}
\fi
\begin{lemma}\label{lem:A1'inside}
Assume that $A_1'\subseteq A_1$,  $A_2'\cap A_1
\ne\emptyset$, and $A_1'\ndom A_2'$. Then, $P_1'$ is memoryless.
\end{lemma}
\ifproofs
\begin{proof}
The lemma follows by applying Lemma~\ref{lem:domination} with the roles
of $\partn$ and $\partnp$ reversed, and observing that $A_1'\cap A_1 =
A_1'$.
\end{proof}
\fi

\begin{lemma}\label{lem:A1'notdom}
Assume that $A_1\ndom A_2$ and $A_1'\subseteq A_1$. If
$A_2'\in\partnp$, and $A_2'\cap A_2 \nonempty$, then $A_1'\ndom A_2'$.
\end{lemma}
\ifproofs
\begin{proof}
We apply Lemma~\ref{lem:ergod}, referring only to the strings $V$ and
$v$ guaranteed by the lemma, and with $B_2=A_2\cap A_2'$. Thus, for any
integer $\MMM$, there exists a string $V\in\partn^\ast$ and a string
$v$ consistent with $V$ such that $\MMM\le\bigstrlen{v[B_2]}
\le\bigstrlen{v[A_2']}$,  while $\nocc{A_1}(V)=0$ and, consequently,
$\bigstrlen{v[A_1']} =0$. Letting $V'=\alphseq[\partnp](v)$, we then
have $\nocc{A_1'}(V')=0$ and $\nocc{A_2'}(V')\ge \MMM$ for arbitrarily
large $\MMM$. Thus, $A_1'\ndom A_2'$.
\end{proof}
\fi

\begin{lemma}\label{lem:Ai'memoryless}
Assume that $A_1$ is not \totdominant, $A_1'\subseteq A_1$, and $P_1'$
is memoryless. Then, for all $a\in A_1'$, $P_1(a|s)$ is independent of
$s\in\states(P_1)$.
\end{lemma}
\ifproofs
\begin{proof}
Since $m>1$ and $A_1$ is not \totdominant, there exists a subalphabet,
say $A_2\in\partn$, such that $A_1\ndom A_2$. Consider a symbol $a\in
A_1'$. Let $s$ be an arbitrary state of $P_1$, and let $U$, $V$, $u$,
and $v$ be the strings guaranteed by Lemma~\ref{lem:ergod} for the
state $s$, with $\MMM =
\max\{\ord(\pswitch),\ord(\pswitchp)\}$. Then, applying~(\ref{eq:IMP}) to
the two IMP representations under consideration, we have
\begin{equation}
  \label{eq:PaUV}
  P(a|uv) = P_1(a|s) \pswitch(A_1|UV) = P_1'(a)\pswitchp(A_1'|U'V'),
\end{equation}
where $U'V'=\alphseq[\partnp](uv)$, and we have relied on the fact that
$P_1'$ is memoryless. Recall from Lemma~\ref{lem:ergod}(\ref{it:UV})
that $\pswitch(A_1|UV)>0$. By our choice of $\MMM$, it follows
from~(\ref{eq:PaUV}) that $P_1(a|s) =
P_1'(a)\pswitchp(A_1'|V')/\pswitch(A_1|V)$, which is independent of
$s$.
\end{proof}
\fi

\begin{lemma}\label{lem:subdivide}
Assume that $A_1$ does not dominate any subalphabet $A_j$, $j>1$, that
shares some $A_\ell'\in\partnp$ with $A_1$. Then, either $P_1$ is
memoryless, or $A_1$ splits into subalphabets in $\partnp$.
\end{lemma}
\begin{proof}
Assume that $A_1$ does not split into subalphabets in $\partnp$. Then,
there exists a subalphabet $A_\ell'\in\partnp$ that intersects $A_1$
but is not contained in it, so $A_1$ shares $A_\ell'$ with some $A_j$,
$j>1$. By the lemma's assumptions, we have $A_1\ndom A_j$. Therefore,
by Lemma~\ref{lem:domination}, $P_1(a|s)$ is independent of
$s\in\states(P_1)$ for all $a\in A_1\cap A_\ell'$. Assume now that
there is also a subalphabet $A_i'\in\partnp$ such that $A_i'\subseteq
A_1$. By Lemma~\ref{lem:A1'notdom}, we have $A_i'\ndom A_\ell'$, and,
therefore, by Lemma~\ref{lem:A1'inside}, $P_i'$ is memoryless. Thus, by
Lemma~\ref{lem:Ai'memoryless}, $P_1(a|s)$ is independent of $s$ also
when $a\in A_i'\subseteq A_1$. Consequently, if $A_1$ does not split in
$\partnp$, since every $a\in A_1$ must belong to some $A_h'\in\partnp$,
and $P_1(a|s)$ is independent of $s\in\states(P_1)$ whether $A_h'$ is
contained in $A_1$ or not, $P_1$ must be memoryless.
\end{proof}

\iffalse
The condition of Lemma~\ref{lem:subdivide} holds trivially when $A_1$
is not dominant, leading to the following result.
\begin{corollary}\label{cor:subdivide}
If $A_1$ is not dominant, then either $P_1$ is memoryless, or $A_1$
splits into subalphabets in $\partnp$.
\end{corollary}
\fi

\begin{lemma}\label{lem:doesnotsubdivide}
Assume that $A_1$ is not totally dominant, and that $A_1$ does not
dominate any subalphabet $A_j$, $j > 1$, that shares some $A_\ell'$
with $A_1$. Then, $P_1$ is memoryless.
\end{lemma}

\begin{proof}
If $P_1$ is not memoryless, then by Lemma~\ref{lem:subdivide}, $A_1$
splits into subalphabets in $\partnp$. Thus, up to re-labeling of
subalphabets, we have $A_1 = A_1'\cup A_2'\cup\cdots\cup A_r'$, where
$A_i'\in\partnp$, $1\le i
\le r \le m'$, with $r>1$. Furthermore, by
Lemma~\ref{lem:Ai'memoryless}, at least one of the $A_i'$, say $A_1'$,
is not memoryless (for, otherwise, $P_1$ would be memoryless). By
Lemma~\ref{lem:A1'inside}, $A_1'$ must dominate all $A_i'$, $2\le i \le
r$, and in particular, $A_1'\dom A_2'$. It follows from this domination
relation that there exists a string $U'\in(\partnp)^\ast$ such that
$\pswitchp(A_2'|U')= 0$, and $\pswitchp(A_1'|U')>0$. By the ergodicity
of $\pswitchp$, we can assume without loss of generality that the
number of occurrences of subalphabets $A_1',A_2',\ldots,A_r'$ in $U'$
is at least $k_1=\ord(P_1)$. Let $u$ be a string consistent with $U'$.
We have $\bigstrlen{u[A_1]}\ge k_1$; let $t\in\states(P_1)$ be the
suffix of length $k_1$ of $u[A_1]$. Consider a symbol $b\in A_2'$, and
let $U''=\alphseq(u)$. Applying~(\ref{eq:IMP}) separately to the two
available IMP representations of $P$, we have
\begin{equation}\label{eq:Pbt=0}
P(b|u) = P_1(b|t)\pswitch(A_1|U'') = P_2'(b|u[A_2'])\pswitchp(A_2'|U') = 0,
\end{equation}
where the last equality follows from our choice of $U'$. On the other
hand, since we also have $\pswitchp(A_1'|U')>0$, we must have
$P(a|u)>0$ for some $a\in A_1'\subseteq A_1$, and, therefore,
$\pswitch(A_1|U'')>0$. Thus, it follows from~(\ref{eq:Pbt=0}) that
$P_1(b|t)=0$. By our assumptions on component processes, there must
also be a state $s\in\states(P_1)$ such that $P_1(b|s)>0$. Since $A_1$
is not totally dominant, there exists a subalphabet, say $A_2$, such
that $A_1\ndom A_2$. Let $B_2=A_2$ and
$\MMM=\max\{\ord(\pswitch),\ord(\pswitchp)\}$. We apply
Lemma~\ref{lem:ergod}(\ref{it:s}), separately to the states $s$ and
$t$, choosing the same string $v$ for both as allowed by the lemma.
Specifically, let $U$ and $V$ be the strings over $\partn$ obtained
from the lemma, and let $\uut$, $\uus $, and $v$ be strings such that
$\uut[A_1]=u't$, $\uus[A_1]=u''s$ for some $u',u''$, both $\uus v$ and
$\uut v$ are consistent with $UV$, and $|v[A_2]|\ge\MMM$. Let
$V'=\alphseq[\partnp](v)$. Clearly, $|V| = |V'|\ge\MMM$, so $V$ and
$V'$ determine states in the respective switches.
Applying~(\ref{eq:IMP}) again, we obtain
\begin{equation}\label{eq:P(b|\uus  v)}
P(b|\uus v) = P_2'(b|\uus[A_2'])\pswitchp(A_2'|V') = P_1(b|s) \pswitch(A_1|V) > 0\,,
\end{equation}
where the last inequality follows from our choice of $s$, and the fact
that $\pswitch(A_1|V) = \pswitch(A_1|UV) > 0$ by our choice of $\MMM$
and by Lemma~\ref{lem:ergod}(\ref{it:UV}). Thus, we must have
$\pswitchp(A_2'|V')>0$. On the other hand, we can also write
\begin{equation}\label{eq:P(b|\uut v)}
P(b|\uut v) = P_2'(b|\uut[A_2'])\pswitchp(A_2'|V') = P_1(b|t) \pswitch(A_1|V) = 0\,,
\end{equation}
where the last equality follows from our choice of $t$. Since, as
previously claimed, $\pswitchp(A_2'|V'){>}0$, it follows from
(\ref{eq:P(b|\uut v)}) that $P_2'(b|\uut[A_2'])=0$, which must hold for
all $b\in A_2'$, a contradiction, since every state of $P_2'$ must have
at least one symbol with positive probability (the argument holds even
if $|\uut [A_2']|<\ord(P_2')$, reasoning with marginal probabilities).
We conclude that $P_1$ must be memoryless.
\end{proof}

The following corollary is an immediate consequence of
Lemma~\ref{lem:doesnotsubdivide}.
\begin{corollary}\label{cor:doesnotsubdivide}
Assume that $A_1$ is not dominant. Then, $P_1$ is memoryless.
\end{corollary}

Assume now that $\pswitch$ is such that no two alphabets in $\partn$
are in mutual domination. As discussed in Section~\ref{sec:domination},
this ensures that $\dom$ defines a strict partial order on $\partn$. We
classify alphabets in $\partn$ into disjoint \emph{layers} $L_i$,
$i{\ge}0$, as follows: Given $L_0,L_1,\ldots,L_{i-1}$, and assuming
that these layers do not exhaust $\partn$, we let $L_i$ consist of the
alphabets that have not been previously assigned to layers, and that
only dominate alphabets contained in layers $L_{i'}$, $0\le i' < i$
(e.g., $L_0$ consists of the non-dominant alphabets in $\partn$). Since
$\partn$ is finite, and every finite set endowed with a strict partial
order has minima, $L_i$ is well defined and non-empty. Thus, for some
$r\ge0$, we can write
\begin{equation}\label{eq:partnlayers}
\partn = L_0 \cup L_1 \cup \cdots \cup L_r\,
\end{equation}
where the layers $L_0,L_1,\ldots,L_r$ are all disjoint and
non-empty.\footnote{The layers $L_i$ correspond to height levels in the
directed acyclic graph associated with the transitive reduction of the
partial order $\dom$.}

We are now ready to present the proofs of Theorems~\ref{th:ambiguities}
and~\ref{th:TD}, which rely on the foregoing lemmas and corollaries,
and on the classification of alphabets into layers $L_i$.

\begin{proof}[Proof of Theorem~\ref{th:ambiguities}]
For the layers in~(\ref{eq:partnlayers}) we prove, by induction on $i$,
that $L_i\subseteq
\partnp$ for $0\le i\le r$. By the definition of $L_0$, alphabets $A_j\in L_0$ are not
dominant. Thus, by Corollary~\ref{cor:doesnotsubdivide}, we must have
$A_j\in\partnp$, since,  by assumption~(\ref{as:noMM}), $A_j$ is not
memoryless. Hence, $L_0\subseteq
\partnp$. Assume now that the induction claim has been proven for
$L_0, L_1,\ldots, L_{i-1}$, $1\le i\le r$. Let $A_j$ be any alphabet in
$L_{i}$. By definition of $L_{i}$, $A_j$ only dominates alphabets in
layers $L_{i'}$, $i'< i$. But, by our induction hypothesis, alphabets
in these layers are elements of $\partnp$, and, thus, they do not share
with other alphabets from $\partn$. Thus, $A_j$ does not dominate any
alphabet $A_h$ with which it shares any $A_\ell'$. By
Lemma~\ref{lem:doesnotsubdivide}, we must have $A_j\in\partnp$, since
$A_j$ is neither totally dominant nor memoryless by the assumptions of
the theorem. Hence, $L_{i}\subseteq
\partnp$, and our claim is proven. Now, it follows
from~(\ref{eq:partnlayers}) that $\partn
\subseteq\partnp$, and, since both $\partn$ and $\partnp$ are
partitions of the same alphabet $\alphabet$, we must have
$\partn=\partnp$.
\end{proof}

\begin{proof}[Proof of Theorem~\ref{th:TD}]
Examining the proof of Theorem~\ref{th:ambiguities}, we observe that
when Condition~(\ref{as:noTD}) is removed, any totally dominant
alphabet must reside in $L_r$, the last layer
in~(\ref{eq:partnlayers}). Furthermore, if there is such an alphabet
$A_{i_0}$, it must be unique, for otherwise there would be alphabets in
mutual domination. Thus, we have $L_r =
\{A_{i_0}\}$, and $A_i\in\partnp$ for all $i\ne i_0$, and, therefore, $A_{i_0}$
splits into the remaining alphabets in $\partnp$ that are not equal to
any $A_i$.
\end{proof}

\subsection{Derivation of Theorem~\ref{theo:memoryless}}\label{app:memoryless}

We start by proving Lemma~\ref{lem:memoryless} of
Subsection~\ref{sec:memoryless}, and then proceed to present an
additional auxiliary lemma, and the proof of
Theorem~\ref{theo:memoryless}.

\begin{proof}[Proof of Lemma~\ref{lem:memoryless}]
Assume $\Ponej$, $j\in\onetwo$, and $\pswitchp$
satisfy~(\ref{eq:P'P''})--(\ref{eq:splitswitch}). We prove that $P(u^n)
= P'(u^n)$ for all lengths $n$ and sequences $u^n\in\alphabet^n$ by
induction on $n$. For $n=0$, the claim is trivially true due to the
convention $P(\lambda) = P'(\lambda) = 1$. Assume that
$P(u^{n-1})=P'(u^{n-1})$ for $n>0$ and all $u^{n-1}\in\alphabet^{n-1}$,
and consider a sequence $u^n=u^{n-1}u_n$. Let $U^n=\alphseq(u^n)$ and
$(U')^n=\alphseq[\partnp](u^n)$, and let $S\in\states(\pswitch)$ and
$S'\in\states(\pswitchp)$ be the states selected by $U^{n-1}$ and
$(U')^{n-1}$, respectively. Clearly, we have $S=\psimap(S')$. By the
definition of $\partnp$, if $U_n=A_i$, $i\in\{2,3,\ldots,m\}$, then
$U'_n = U_n$, and we have
\begin{eqnarray}
P'(u^n)&=&P'(u^{n-1})P'(u_n|u^{n-1})
=
P'(u^{n-1})\pswitchp(A_i|S')P_i(u_n\left|u^{n-1}[A_i])\right.)\nonumber\\
&=&
P(u^{n-1})\pswitch(A_i|S)P_i(u_n\left|u^{n-1}[A_i])\right.)
= P(u^n)\,,\label{eq:P'=P for A_i}
\end{eqnarray}
where the second and last equalities follow from the definitions of the
respective IMPs, and the third equality follows from the induction
hypothesis and~(\ref{eq:splitswitch}).
 On the other hand, if $U_n=A_1$, then
 $U_n'= B_j$ for some $j\in\onetwo$, and we have
\begin{eqnarray}
P'(u^n)&=&P'(u^{n-1})P'(u_n|u^{n-1})
= P'(u^{n-1})\pswitchp(B_j|S')\Ponej(u_n)\nonumber\\
&=&
P(u^{n-1})\pswitch(A_1|S)P_1(B_j)\frac{P_1(u_n)}{P_1(B_j)}
= P(u^n)\,,\label{eq:P'=P for A_1}
\end{eqnarray}
where, this time, the third equality follows from the induction
hypothesis, (\ref{eq:splitswitch}), and~(\ref{eq:P'P''}) (we recall
that $P_1$, $\Ponej[1]$, and $\Ponej[2]$ are memoryless).
 This completes the induction proof and establishes
that $P'=P$.

To prove the ``only if'' part of the lemma, we assume that $P'=P$, and
consider a sufficiently long, arbitrary string $u^n$ such that
$P(u^n)>0$. Let $U'=\alphseq[\partnp](u^{n-1})$, and assume first that
$u_n\in A_i$ for some $i\ge 2$. Then, similarly to~(\ref{eq:P'=P for
A_i}) (but proceeding from the inside out), and noting that
$\alphseq(u^{n-1})=\psimap(U')$, we can write
\begin{eqnarray}
P'(u^{n-1})\pswitchp(A_i|U')P_i(u_n\left|u^{n-1}[A_i])\right.) &=& P'(u^n) = P(u^n)\nonumber\\
&=&
P(u^{n-1})\pswitch(A_i|\psimap(U'))P_i(u_n\left|u^{n-1}[A_i])\right.)\,.\label{eq:P'=P for A_i converse}
\end{eqnarray}
Since $P'=P$, and $P(u^n)>0$,~(\ref{eq:P'=P for A_i converse}) can be
simplified to
\begin{equation}\label{eq:Pw=Pw'}
\pswitchp(A_i|U') = \pswitch(A_i|\psimap(U')),\quad i \in \twom\,,
\end{equation}
for arbitrary $U'\in (\partnp)^{n-1}$ of positive probability. Consider
now the case $u_n = b\in B_j$, $j \in \onetwo$. Then, in analogy
with~(\ref{eq:P'=P for A_1}), we write
\begin{equation}
P'(u^{n-1})\pswitchp(B_j|U')\Ponej(b) = P'(u^n) = P(u^n)
=
P(u^{n-1})\pswitch(A_1|\psimap(U'))P_1(b)\,.\label{eq:P'=P for A_1 converse}
\end{equation}
Adding over all $b\in B_j$ and simplifying, we obtain
\begin{equation}\label{eq:Pw=Pw'Bj}
\pswitchp(B_j|U') = \pswitch(A_1|\psimap(U'))P_1(B_j),\quad j\in\onetwo\,,
\end{equation}
again for arbitrary $U'$.
Conditions~(\ref{eq:states'})--(\ref{eq:splitswitch}) now follow
readily from~(\ref{eq:Pw=Pw'}) and~(\ref{eq:Pw=Pw'Bj}) (which imply, in
particular, that $\kswitch=\kswitch'$), and Condition~(\ref{eq:P'P''})
follows by substituting the right-hand side of~(\ref{eq:Pw=Pw'Bj}) for
$\pswitchp(B_j|U')$ in~(\ref{eq:P'=P for A_1 converse}) and solving for
$\Ponej(b)$.
\end{proof}

We say that the representations $\IPi$ and $\IPip$ of an IMP $P$
\emph{coincide up to memoryless components} if the set of component
processes of positive order is the same in both representations. The
following lemma establishes the uniqueness of canonical partitions.
\begin{lemma}\label{lem:samecanonical}
Let $\IPi$ and $\IPip$ be IMP representations of a process $P$ that
coincide up to memoryless components, and such that both are canonical.
Then, $\partn = \partnp$.
\end{lemma}
\begin{proof}
Assume that $\partn\ne\partnp$, and let $\partnpp$ be the smallest
common refinement of $\partn$ and $\partnp$ (i.e., $\partnpp =
\left\{A_i\cap A_j'\,\left|\,A_i\in\partn,
\;A_j'\in\partnp, \;A_i\cap A_j'\ne\emptyset\right.\right\}$).
By repeated application of Lemma~\ref{lem:memoryless}, there exists an
IMP representation
$\intlv_{\partnpp}(P_1'',P_2'',\ldots,P_{m''}'';\pswitchpp)$ of $P$.
This representation is a memoryless refinement of both $\IPi$ and
$\IPip$. Since $\partn\ne\partnp$, there exists an alphabet, say
$A_1'\in\partnp$ such that $A_1'\not\in\partn$, $P_1'$ is memoryless,
and we can assume without loss of generality that $A_1'$ intersects at
least two alphabets, $A_1$ and $A_2$, in $\partn$ (otherwise, we can
switch the roles of $\partn$ and $\partnp$). Let $B_1=A_1'\cap A_1$ and
$B_2=A_1'\cap A_2$, so that $B_1,B_2\in\partnpp$. Applying
Lemma~\ref{lem:memoryless} separately to $\IPi$ and to $\IPip$ with
respect to the refinement $\intlv_{\partnpp}$, we can write, for any
$S''\in\states(\pswitchpp)$, and denoting
$S=\psimap[\partn,\partnpp](S'')$ and
$S'=\psimap[\partnp,\partnpp](S'')$,
\begin{equation*}
\pswitchpp(B_1|S'') = \pswitch\left(\left.A_1\right|S\right)P_1(B_1)
=\pswitchp\left(\left.A_1'\right|S'\right)P_1'(B_1),
\end{equation*}
where $P_1(B_1)$ and $P_1'(B_1)$ are nonzero. (Notice that the equation
holds also when $B_1=A_1$, i.e., when $A_1$ is not actually refined in
$\partnpp$.) Therefore, we can write
\begin{equation}\label{eq:B_1'}
\pswitch\left(\left.A_1\right|S\right) =
\frac{\pswitchp\left(\left.A_1'\right|S'\right)P_1'(B_1)}
{P_1(B_1)}\,.
\end{equation}
 Using a similar argument for $B_2$ and $A_2$, we obtain
\begin{equation}\label{eq:B_2'}
\pswitch\left(\left.A_2\right|S\right) =
\frac{\pswitchp\left(\left.A_1'\right|S'\right)P_1'(B_2)}
{P_1(B_2)}\,.
\end{equation}
It follows from~(\ref{eq:B_1'}) and~(\ref{eq:B_2'}) that if
$\pswitchp\left(\left. A_1'\right|S'\right)=0$, then $\pswitch(A_1|S)
=\pswitch(A_2|S) = 0$, and, otherwise,
\[
\frac{\pswitch\left(A_2|S\right)}
{\pswitch\left(A_1|S\right)} =
\frac{P_1(B_1)P_1'(B_2)}{P_1'(B_1)P_1(B_2)}
\defined
\gamma,
\]
where $\gamma>0$ is independent of $S''$ (and of $S$). Observing that
$S$ can assume any value in $\states(\pswitch)$, we conclude, by
Lemma~\ref{lem:memoryless} and the remarks following its statement,
that $A_1$ could be merged with $A_2$, contradicting the assumption
that $\IPi$ is canonical. Thus, we must have $\partn =\partnp$.
\end{proof}

\begin{proof}[Proof of Theorem~\ref{theo:memoryless}]
Assume $P=P'$. Since there are no dominant alphabets in either
representation, it follows from Corollary~\ref{cor:doesnotsubdivide}
that the representations must coincide up to memoryless components. It
then follows from Lemma~\ref{lem:samecanonical} that the canonical
partitions of $\IPi$ and $\IPip$ must be identical, and, thus, since
they generate the same process, we must have
$(\IPi)^\ast\equiv(\IPip)^\ast$. The ``if'' part is straightforward,
since $(\IPi)^\ast$ generates $P$, and $(\IPip)^\ast$ generates $P'$.
\end{proof}

\section{The deinterleaving scheme: derivations}
\label{app:consistency}
We will prove
Theorem~\ref{th:consistency} through the auxiliary
Lemmas~\ref{lem:mismatch} and~\ref{lem:canonicalwins} below, for which
we need some additional definitions.

Let $F=(S,s_0,f)$ be an FSM, and let $P$ and $Q$ be processes generated
by $F$, such that $P$ is ergodic. The divergence (relative to $F$)
between $P$ and $Q$ is defined as
\begin{equation}\label{eq:DPQ}
D(P||Q) = \sum_{s\in S} P(s)\bigdiv{P(\cdot|s)}{Q(\cdot|s)}\,,
\end{equation}
where $P(s)$ denotes the stationary probability of the state $s\in S$,
and $\bigdiv{P(\cdot|s)}{Q(\cdot|s)}$ denotes the Kullback-Leibler
divergence between the conditional distributions $P(\cdot|s)$ and
$Q(\cdot|s)$. It is well known (see, e.g.,~\cite{CCC'87}) that
$D(P||Q)$ as defined in~(\ref{eq:DPQ}) is equal to the asymptotic
normalized Kullbak-Liebler divergence between the processes $P$ and
$Q$, namely,
\[
D(P||Q) = \lim_{n\to\infty}\frac{1}{n}\sum_{z^n\in\alphabet^n}P(z^n)\log\frac{P(z^n)}{Q(z^n)}\,.
\]

Let $\VFP$ denote the set of parameter vectors corresponding to ergodic
\emph{unconstrained} FSMSs based on $\FP$, and let $\VFPb$ denote its topological
closure. Assuming full parametrization, this set is a convex polytope
in $\dpartn$-dimensional Euclidean space. The boundary of $\VFPb$
consists of parameter vectors with certain transition probabilities set
to zero or one. Some of these vectors do not correspond to ergodic
FSMS, namely, those that make some of the marginal probabilities of
states in $S$ vanish (e.g., parameter vectors where the probabilities
of all the transitions leading to a state vanish). Let $\VFPI$, in
turn, denote the set of parameter vectors of IMP-constrained FSMSs
based on $\FP$, and $\VFPIb$ its topological closure. The set $\VFPIb$
is a closed $\dpartni$-dimensional hypersurface within $\VFPb$,
determined by the parameter relations implicit in~(\ref{eq:PF}). As
before, boundary points in $\VFPIb$ are either in $\VFPI$, or do not
correspond to valid IMPs. We shall make use of these relations in the
sequel.

The following lemma will be useful in proving the first claim of
Theorem~\ref{th:consistency}.
\begin{lemma}
\label{lem:mismatch} Let
$P=\IPi(P_1,P_2,\ldots,P_m;\pswitch)$,  and let
$\bldk=(k_1,k_2,\ldots,k_m,k_\swltr)$ be the corresponding order
vector. Let $\partn'$ be a partition of $\alphabet$ incompatible with
$P$, and $\bldk'$ an arbitrary order vector of dimension $|\partnp|+1$.
Then, for a sample $z^n$ from $P$, and for any $\beta\ge 0$, we have
\[
\kcost[\partnp,\bldk'](z^n) > \kcost[\partn,\bldk](z^n)\quad\text{a.s.\ \   as}\; n\to\infty\,.
\]
\end{lemma}
\begin{proof}
Let $\refF$ be a
common refinement\footnote{ %
It is always possible to construct a common refinement of two FSMs,
e.g., one whose state set is the Cartesian product of the state sets of
the refined FSMs.} %
 of $F=\FP$ and $F'=\FPkp$.
Let $V=\VFP[\refF]$ denote the space of all valid parameter vectors for
FSM sources based on $\refF$, and let $\VFPb[\refF]$ denote its
topological closure. The constraints satisfied by
\ISMP\ sources based on $F$ and $F'$ are
extended to their representations in $V$ (notice that a refinement
increases the dimension of the parameter vector by ``cloning''
parameters, together with their constraints). Thus, as mentioned in the
discussion immediately preceding the lemma, the set of all
IMP-constrained FSMSs based on $F'$ maps to a lower-dimensional
hypersurface $V'=\VFPI[\refF]\subseteq V$, with closure $\Vclosed'$. We
claim that the representation of $P$ in $V$ is outside the closed
hypersurface $\Vclosed'$, and, thus, at positive Euclidean (or $L_1$)
distance from it. To prove the claim, we first notice that since
$\partnp$ is, by assumption, incompatible with $P$, no valid
IMP-constrained assignment of parameters for $F'$ can generate $P$,
and, thus, $P\not\in V'$. Furthermore, since points in
$\Vclosed'\setminus V'$ correspond to ``invalid'' IMPs with unreachable
states, we must have  $P
\not\in\Vclosed'$, and, therefore, $P$ is at positive distance from
$\Vclosed'$, as claimed. The ergodicity of $P$ also implies that, in
its representation in $V$, all the states of $\refF$ have positive
stationary probabilities. Applying Pinsker's inequality on a state by
state basis in~(\ref{eq:DPQ}) for $\refF$, we conclude that for any
process $P'\in V'$, we have
\begin{equation}\label{eq:divergence}
D(P||P') \ge \Delta \,,
\end{equation}
for some constant $\Delta > 0$. Now, recall that $\pmlF[\refF](z^n)$
denotes the ML probability of $z^n$ with respect to $\refF$ for an
\emph{unconstrained} FSMS.
It follows from the definition of $\pmlF[\refF](z^n)$ and of the
divergence $D(\cdot||\cdot)$ in~(\ref{eq:DPQ}) that for any process $Q$
generated by $\refF$, we have
\begin{equation}\label{eq:logQ}
-\log Q(z^n) = -\log\pmlF[\refF](z^n) + n \bigdiv{\pmlF[\refF]}{Q}\,.
\end{equation}
In particular, since $\refF$ can generate any process that either $F$
or $F'$ can generate, it can assign to $z^n$ its IMP-constrained ML
probabilities with respect to $F$ and $F'$ which are, respectively,
$\hpF(z^n)=2^{-\hhF(z^n)}$ and $\hpFp(z^n)=2^{-\hhFp(z^n)}$.
Applying~(\ref{eq:logQ}) to $Q=\hpF$ and $Q=\hpFp$ separately,
subtracting on each side of the resulting equations, and dividing by
$n$, we obtain
\begin{equation}\label{eq:deltaH}
\frac{1}{n}\left(\hhFp(z^n)-\hhF(z^n)\right) = \bigdiv{\pmlF[\refF]}{\hpFp}-\bigdiv{\pmlF[\refF]}{\hpF}\,.
\end{equation}
Now, since $z^n$ is a sample from $P$, the empirical measures
$\pmlF[\refF]$ and $\hpF$ tend to the true process $P$ almost surely in
the divergence sense, i.e., $\bigdiv{\pmlF[\refF]}{P}\to 0$ and
$\bigdiv{\hpF}{P}\to 0$ a.s. as $n\to\infty$. Also, an empirical
conditional probability value in either $\pmlF[\refF]$ or $\hpF$ is
surely  zero if the corresponding parameter in $P$ is zero, and almost
surely bounded away from zero otherwise. Hence, we also have
$\bigdiv{\pmlF[\refF]}{\hpF}\to 0$ a.s.\ as $n\to\infty$. On the other
hand, since $\hpFp\in V'$,~(\ref{eq:divergence}) applies with
$P'=\hpFp$,  so we have $\bigdiv{P}{\hpFp}\ge \Delta > 0$, and, using a
similar convergence argument, $\bigdiv{\pmlF[\refF]}{\hpFp}\ge
\Delta> 0$ a.s.\ as $n\to\infty$. Thus, it follows from~(\ref{eq:deltaH})
that
\[
\frac{1}{n}\left(\hhFp(z^n)-\hhF(z^n)\right)  \ge \Delta > 0\quad\text{a.s. as }n\to\infty,
\]
which implies, by~(\ref{eq:costpartn}),
\begin{equation}\label{eq:Delta}
\frac{1}{n}\Bigl(\kcost[\partnp,\bldk'](z^n) - \kcost[\partn,\bldk](z^n)\Bigr) \ge \Delta > 0\,
\;\,\mbox{a.s.\ as } n{\to}\infty\,,
\end{equation}
since the contribution of the $O(\log n)$ penalty terms to the costs
vanishes asymptotically in this case, for any choice of $\beta \ge 0$.
\end{proof}

The following lemma, in turn, will be useful in establishing the second
claim of Theorem~\ref{th:consistency}.
\begin{lemma}\label{lem:canonicalwins}
Let $\partn$, $\partnp$, $\IPi$ and $\IPip$ be as defined in
Lemma~\ref{lem:memoryless}, so that $\IPip$ is a memoryless refinement
of $\IPi$. Let $\bldk=(0,k_2,\ldots,k_m,k_\swltr)$ be the order vector
corresponding to $\IPi$, and $\bldk'=(0,0,k_2,\ldots,k_m,k_\swltr)$
that of $\IPip$. For a sample $z^n$ from $P$, and an appropriate choice
of $\beta$, we have: if $k_\swltr
>0$, then
\begin{equation}\label{eq:aswin}
\kcost[\partnp,\bldk'](z^n) > \kcost[\partn,\bldk](z^n)\quad\text{a.s.\ \   as}\; n\to\infty\,,
\end{equation}
while if $k_\swltr=0$, then
\begin{equation}\label{eq:tie}
\kcost[\partnp,\bldk'](z^n) = \kcost[\partn,\bldk](z^n)\,.
\end{equation}
\end{lemma}
\begin{proof}
We first notice that, by Lemma~\ref{lem:memoryless},
$\mathcal{P}_{\mathcal{I}} (\mathcal{F}_{\Pi,\kvec})$ can alternatively
be characterized as the subset of
$\mathcal{P}_{\mathcal{I}}(\mathcal{F}_{\Pi',\kvec'})$ formed by
distributions such that the switch process $\pswitchp$ satisfies the
following two constraints, where $\psimap$ denotes the mapping defined
prior to Lemma~\ref{lem:memoryless}:
\begin{enumerate}
\renewcommand{\theenumi}{\alph{enumi}}
\item\label{refine1}
If $S',S'' \in \states(\pswitchp)$ satisfy $\Psi(S')=\Psi(S'')$
then the corresponding conditional distributions coincide;
\item\label{refine2}
For every $S \in \states(\pswitchp)$, $\pswitchp(B_2|S) = \gamma
\pswitchp(B_1|S)$ for some parameter $\gamma$, independent of $S$.
\end{enumerate}
Clearly, the dimension of both parametrizations remains
$\kappa(\Pi,\kvec)$. It then follows from the definition of empirical
entropy of an IMP and from~(\ref{eq:defineML}) that
\begin{equation}
\label{eq:constemp}
\hat{H}_{\Pi,\kvec} (z^n) = \hat{H}_0 (z^n[B_1]) + \hat{H}_0 (z^n[B_2]) +
\sum_{i=2}^m \hat{H}_{k_i} (\bldz_i) - \log \tilde{P}'_\swltr (\alphseq[\partn'](z^n))
\end{equation}
where $\tilde{P}'_\swltr(\alphseq[\partn'](z^n))$ denotes the ML
probability, subject to the above two constraints, of the switch
sequence $\alphseq[\partn'](z^n)$. Therefore,
\begin{equation}
\label{eq:diffemp}
\hat{H}_{\Pi,\kvec} (z^n) - \hat{H}_{\Pi',\kvec'} (z^n) =
- \log \tilde{P}'_\swltr (\alphseq[\partn'](z^n)) - \hat{H}_{k_\swltr} (\alphseq[\partn'](z^n))
\end{equation}
which depends on $z^n$ only through $\alphseq[\partn'](z^n)$. The above
difference is obviously nonnegative, since $\Pi'$ is a refinement of
$\Pi$; equivalently, looking at the right-hand side of
(\ref{eq:diffemp}), the maximization leading to $\tilde{P}'_\swltr
(\alphseq[\partn'](z^n))$ involves more constraints than the one
leading to $\hat{H}_{k_\swltr} (\alphseq[\partn'](z^n))$. Recalling the
difference in model sizes computed in~(\ref{eq:comparekappa}), we
obtain, together with~(\ref{eq:diffemp}), that
\begin{eqnarray}
C_{\Pi',\kvec'} (z^n) - C_{\Pi,\kvec} (z^n) &=& \hat{H}_{k_\swltr} (\alphseq[\partn'](z^n))
+ \beta m(m+1)^{k_\swltr} \log (n+1) \nonumber \\
&-& [ - \log \tilde{P}'_\swltr (\alphseq[\partn'](z^n)) + \beta ((m-1)m^{k_\swltr} + 1) \log (n+1) ] \,.
\label{eq:costdiff}
\end{eqnarray}
Thus, the left-hand side of~(\ref{eq:costdiff}) is equal to the
difference between penalized ML probabilities for a switch sequence of
length $n$ on $\Pi'$, for two candidate models. The first model is
Markov of order $k_\swltr$, whereas the second model differs from the
plain Markov one in that states of $(\Pi')^{k_\swltr}$ have merged
according to the mapping $\Psi$, so that the number of states is now
$m^{k_\swltr}$ (constraint~(\ref{refine1}) above), and imposes the
additional constraint~(\ref{refine2}) on the conditional probabilities
of $B_1$ and $B_2$ (notice that the number of free parameters in this
model is indeed $(m-1)m^{k_\swltr} + 1$). Since, by our assumptions,
the number of states of the underlying switch process is $m^{k_\swltr}$
and the process does satisfy the additional constraint~(\ref{refine2}),
the left-hand side of~(\ref{eq:costdiff}) can be viewed as a penalized
ML test of two models, the minimal, ``true'' one, and a refinement of
it. When $k_\swltr = 0$, the refinement is trivial and the penalty
difference is $0$, implying (\ref{eq:tie}). When $k_\swltr > 0$, our
analysis, presented next, will rely on tools developed in~\cite{WF94}
to study refinements of the type given by constraint~(\ref{refine1}),
which will be extended here to deal also with the type of refinement
given by constraint~(\ref{refine2}). As in~\cite{WF94}, we will show
the strong consistency of the penalized ML test for suitable~$\beta$.

Specifically, given a sequence $Z^n$ over $(\Pi')^n$, we start by
defining the following ``semi-ML'' Markov probability distribution
$\chubchik{P}'_\swltr$ of order $k_\swltr$: For every $S \in
(\Pi')^{k_\swltr}$ and $i=2,\cdots,m$, we define $\chubchik{P}'_\swltr
(A_i|S) = P_\swltr (A_i|S)$ if $S \in
\Pi^{k_\swltr}$ (i.e., $S$ is a $k_\swltr$-tuple over $(\Pi')^{k_\swltr}$
not containing either $B_1$ or $B_2$, and is therefore an unrefined
state of $\Pi^{k_\swltr}$), and $\chubchik{P}'_\swltr (A_i|S) =
\hat{P}'_\swltr (A_i|\Psi(S))$ otherwise, where $\hat{P}'_\swltr
(A_i|\bar{S})$ denotes the ratio between the number of occurrences of
$A_i$ following a state $\bar{S}$ in $Z^n$, and the number of
occurrences of $\bar{S}$, where $\bar{S}$ can be either in
$\Pi^{k_\swltr}$ (as is $\Psi(S)$ in this case) or, more generally, in
$(\Pi')^{k_\swltr}$. The distribution is completely determined by
further setting, for every $S \in (\Pi')^{k_\swltr}$, the relation
$\chubchik{P}'_\swltr (B_2|S) =
\hat{\gamma} \chubchik{P}'_\swltr (B_1|S)$, where
\[
\hat{\gamma} \defined \frac{N_{B_2}(Z^n)}{N_{B_1}(Z^n)}
\]
is the ML estimate of $\gamma$ based on $Z^n$, given by the ratio
between the number of occurrences of $B_2$ and $B_1$ in $Z^n$
(independent of $S$), provided $N_{B_1}(Z^n) > 0$. Otherwise, if
$N_{B_1}(Z^n) = 0$, we let $\chubchik{P}'_\swltr (B_1|S) = 0$. Notice
that $\hat{P}'_\swltr (A_i|S)$ is the ML estimate of $P'_\swltr
(A_i|S)$ regardless of the constraint relating $P'_\swltr(B_2|S)$ and
$P'_\swltr(B_1|S)$. Since, in order to obtain the (constrained) ML
probability $\tilde{P}'_\swltr (Z^n)$, one can first maximize over
$\gamma$ and then perform independent maximizations of the conditional
probabilities for each state, it is easy to see that, for any $Z^n \in
(\Pi')^n$, we have
\begin{equation}
\label{eq:semibound}
P'_\swltr (Z^n) \le \chubchik{P}'_\swltr (Z^n) \le \tilde{P}'_\swltr (Z^n)
\end{equation}
justifying our reference to $\chubchik{P}'_\swltr$ as a ``semi-ML''
Markov probability distribution.

Another (non-constrained) ``semi-ML'' Markov probability distribution
$\kumkum{P}'_\swltr$ of order $k_\swltr$ is defined as follows: For
every $S \in (\Pi')^{k_\swltr} \cap
\Pi^{k_\swltr}$ we define ${\kumkum{P}'_\swltr} (A_i|S) = P_\swltr
(A_i|S)$, $i=2,\cdots,m$, and ${\kumkum{P}'_\swltr} (B_2|S) =
\hat{\gamma}_S {\kumkum{P}'_\swltr} (B_1|S)$, where $\hat{\gamma}_S$
denotes the ratio between the number of occurrences of $B_2$ and $B_1$
following state $S$ in $Z^n$, provided the latter number is positive
(otherwise, we let ${\kumkum{P}'_\swltr} (B_1|S) = 0$). For all other
states $S \in (\Pi')^{k_\swltr}$ and every $Z \in \Pi'$, we define
${\kumkum{P}'_\swltr} (Z|S) = \hat{P}'_\swltr (Z|S)$.

Notice that for states in $(\Pi')^{k_\swltr} \cap \Pi^{k_\swltr}$,
${\kumkum{P}'_\swltr}$ differs from $\chubchik{P}'_\swltr$ in that the
ratio between the conditional probabilities of $B_2$ and $B_1$ depends
on $S$ (while the conditional probabilities of all $A_i$,
$i=2,\cdots,m$, under the two measures, coincide, and are independent
of $Z^n$). For the other states, both ${\kumkum{P}'_\swltr}$ and
$\chubchik{P}'_\swltr$ use ML estimates (which are constrained for the
latter distribution). The key observation is then that
\begin{equation}
\label{eq:key}
- \log \tilde{P}'_\swltr (\alphseq[\partn'](z^n)) - \hat{H}_{k_\swltr} (\alphseq[\partn'](z^n))
= - \log \chubchik{P}'_\swltr (\alphseq[\partn'](z^n)) + \log {\kumkum{P}'_\swltr}
(\alphseq[\partn'](z^n)) \,.
\end{equation}
Now, the probability $P_{\rm{err}}(n)$ of the error event is given by
\begin{equation}
\label{eq:error}
P_{\rm{err}}(n) \defined \sum_{z^n: C_{\Pi',\kvec'} (z^n) \le
C_{\Pi,\kvec} (z^n)} P (z^n) = \sum_{Z^n \in \mathcal{E}} P'_\swltr (Z^n)
\end{equation}
where $\mathcal{E}$ denotes the subset of switch sequences $Z^n$ over
$(\Pi')^n$ satisfying
\[
\hat{H}_{k_\swltr} (Z^n) + \beta m(m+1)^{k_\swltr} \log (n+1)
\le - \log \tilde{P}'_\swltr (Z^n) + \beta [(m-1)m^{k_\swltr} + 1] \log (n+1)
\]
and the second equality in~(\ref{eq:error}) follows
from~(\ref{eq:costdiff}). By~(\ref{eq:key}), $Z^n \in \mathcal{E}$ if
and only if
\[
- \log \chubchik{P}'_\swltr (Z^n) \ge - \log {\kumkum{P}'_\swltr} (Z^n) +
\beta [m(m+1)^{k_\swltr}-(m-1)m^{k_\swltr}-1] \log (n+1)
\]
or, equivalently,
\[
\chubchik{P}'_\swltr (Z^n) \le (n+1)^{-\beta [m(m+1)^{k_\swltr}-(m-1)m^{k_\swltr}-1]}
{\kumkum{P}'_\swltr} (Z^n) \,.
\]
Therefore, by the first inequality in~(\ref{eq:semibound}), the
rightmost summation in~(\ref{eq:error}) can be upper-bounded to obtain
\begin{equation}
\label{eq:mainbound}
P_{\rm{err}}(n) \le (n+1)^{-\beta [m(m+1)^{k_\swltr}-(m-1)m^{k_\swltr}-1]}
\sum_{Z^n \in (\Pi')^n} {\kumkum{P}'_\swltr} (Z^n) \,.
\end{equation}
Notice that the probability distributions in the summation in the
right-hand side of (\ref{eq:mainbound}) depend on $Z^n$. Clearly, when
restricted to sequences $Z^n$ giving rise to the same distribution, the
partial sum is upper-bounded by $1$. Therefore, the overall sum is
upper-bounded by the number $N$ of distinct such distributions. Now,
there are $(m+1)^{k_\swltr} - (m-1)^{k_\swltr}$ states given by
$k_\swltr$-tuples containing either $B_1$ or $B_2$ and, by the
definition of ${\kumkum{P}'_\swltr}$, for each of these states there
are at most $(n+1)^{m+1}$ possible conditional distributions, given by
the composition of the corresponding substring in $Z^n$. For each of
the remaining $(m-1)^{k_\swltr}$ states, the definition of
${\kumkum{P}'_\swltr}$ implies that there are at most $(n+1)^2$
possible conditional distributions. Therefore,
\[
N \le (n+1)^{2(m-1)^{k_\swltr} + [(m+1)^{k_\swltr} - (m-1)^{k_\swltr}] (m+1)}
\]
implying
\begin{equation}
\label{eq:borel}
P_{\rm{err}}(n) \le (n+1)^{2(m-1)^{k_\swltr} + [(m+1)^{k_\swltr} - (m-1)^{k_\swltr}] (m+1)
- \beta [m(m+1)^{k_\swltr}-(m-1)m^{k_\swltr}-1]} \,.
\end{equation}
Since $m \ge 2$ and $k_\swltr \ge 1$, it can be readily shown that, for
any $\beta > 3$, the exponent in the right-hand side
of~(\ref{eq:borel}) is less than $-1$. Thus, $P_{\rm{err}}(n)$ is
summable and the result follows from the Borel-Cantelli lemma.
\end{proof}

With these tools in hand, we are now ready to prove
Theorem~\ref{th:consistency}.

\begin{proof}[Proof of Theorem~\ref{th:consistency}]
Define the set
\[
\sspace' = \left\{\,(\partnp,\bldk')\;|\;\partnp\;\text{ is \emph{incompatible} with }\; P\,\right\}\,.
\]
To establish the first claim of the theorem, we will prove that
$\left(\hat{\partn}(z^n),\hat{\bldk}(z^n)\right)\not\in\sspace'$
a.s.~as $n\to\infty$. Consider a partition $\partnbar$ compatible with
$P$, denote by $\bldkbar$ the associated order vector, and let
$\bar{\dpartni}=\dpartni(\partnbar,\bldkbar)$. Let
$\dpartni_0>\bar{\dpartni}$ denote a threshold for model sizes, which
is independent of $n$, and will be specified in more detail later on.
Write $\sspace' =
\sspace_1 \cup \sspace_2$, where
\[
\sspace_1 = \left\{\,(\partnp,\bldk')\in\sspace'\;|\;\dpartni(\partnp,\bldk') <
\dpartni_0\,\right\}\,,
\]
and $\sspace_2 = \sspace'\setminus\sspace_1$. Clearly, $\sspace_1$ is
finite and its size is independent of $n$. By Lemma~\ref{lem:mismatch},
for each pair $(\partnp,\bldk')\in\sspace_1$, we have
$\kcost[\partnp,\bldk'](z^n)>
\kcost[\partnbar,\bldkbar](z^n)$ a.s.~as $n\to\infty$, for any penalization
coefficient $\beta \ge 0$. Thus, the search in~(\ref{eq:rulePi}),
almost surely, will not return a pair from $\sspace_1$. It remains to
prove that it will not return a pair from $\sspace_2$ either. As
mentioned, the difficulty here is that the size of $\sspace_2$ (and of
the IMP models associated with pairs in $\sspace_2$) is not bounded as
$n\to\infty$, and we cannot establish the desired result with a finite
number of applications of Lemma~\ref{lem:mismatch}. As before, we adapt
some tools from~\cite{WF94} to IMP-constrained FSMSs.

For $(\partnp,\bldk')\in\sspace_2$, let $\Pbadp$ denote the probability
that a solution with $(\partnp,\bldk')$ is preferred over
$(\partnbar,\bldkbar)$ in the minimization. Define
\[
\badset =
\left\{ z^n \,|\,
\kcost[\partnp,\bldk'](z^n)\le\kcost[\partnbar,\bldkbar](z^n)\,\right\}\,.
\]
Clearly, we have
\begin{equation}\label{eq:Pbadset}
\Pbadp \le \sum_{z^n\in \badset}P(z^n)\,.
\end{equation}
By the definitions of $\badset$ and of the cost function
in~(\ref{eq:costpartn}), and denoting
$\dpartni'=\dpartni(\partnp,\bldk')$, we have, for $z^n\in\badset$,
\begin{equation}\label{eq:Hbound}
\hhF[\partnbar,\bldkbar](z^n) \ge \hhF[\partnp,\bldk'](z^n)+
\beta(\dpartni'-\dpartnib)\log(n+1)\,.
\end{equation}
Recalling that  $P(z^n) \le \hpF[\partnbar,\bldkbar](z^n)$
by~(\ref{eq:defineML}), and that $\hhF[\partnp,\bldk'](z^n)=-\log
\hpF[\partnp,\bldk'](z^n)$, it follows
from~(\ref{eq:Hbound}) that
\[
P(z^n) \le
(n+1)^{\beta(\dpartnib-\dpartni')} \hpF[\partnp,\bldk'](z^n)
 \,,\quad  z^n\in\badset\,,
\]
and, hence, together with~(\ref{eq:Pbadset}), and applying an obvious
bound, we obtain
\begin{equation}\label{eq:sumH}
\Pbadp \le (n+1)^{\beta(\dpartnib-\dpartni')}
\sum_{z^n\in\badset} \hpF[\partnp,\bldk'](z^n)
\le
(n+1)^{\beta(\dpartnib-\dpartni')}
\sum_{z^n\in\alpF^n} \hpF[\partnp,\bldk'](z^n)\,.
\end{equation}
In analogy to the reasoning following~(\ref{eq:mainbound}) in the proof
of Lemma~\ref{lem:canonicalwins}, the summation on the right-hand side
of~(\ref{eq:sumH}) can be upper-bounded by the number of different
empirical distributions (or
\emph{types}) for IMPs based on $\FPkp$ and sequences of length $n$. It
is well established (see, e.g.,~\cite{MSW-isit07}) that
$(\alpha_i-1)\alpha_i^{k_i}$ counts suffice to determine the empirical
distribution for the Markov component $P_i$ (and similarly for the
switch $\pswitch$). Hence, recalling~(\ref{eq:IMPparams}), we conclude
that $\dpartni'=\dpartni(\partnp,\bldk')$  counts suffice to determine
an empirical distribution $\hpF[\partnp,\bldk'](z^n)$, and, therefore,
the number of such distributions is upper-bounded (quite loosely) by
$(n+1)^{\dpartni'}$.
 Thus, it follows from~(\ref{eq:sumH}) that
\begin{equation}\label{eq:Pbad}
\Pbadp \le (n+1)^{\beta(\dpartnib-\dpartni') + \dpartni'}\,.
\end{equation}
We next bound the number of pairs $(\partnp,\bldk')$ satisfying
$\dpartni(\partnp,\bldk')= \dpartni'$ for a given
$\dpartni'\ge\dpartni_0$. The number of partitions $\partnp$ is
upper-bounded by $\alpha^\alpha$, where $\alpha=|\alpF|$. For a given
partition, with, say $|\partnp|=m$, we need an assignment of process
orders $k_i'$, $i\in\onemw$. If $|A_i'|=1$, the only valid assignment
is $k_i'=0$, while if $|A_i'|\ge 2$, we must have $k_i'\le
\log \dpartni'$. Thus, since $m\le \alpha$,
the number of pairs sought is upper-bounded by $\alpha^\alpha (\log
\dpartni')^{\alpha+1}$. We notice also that, for $z^n\in\badset$ and
sufficiently large $n$, we must have $\dpartni'
\le n$ (actually, $\dpartni'=o(n)$), for otherwise the penalty component of
$\kcost[\partnp,\bldk'](z^n)$ on its own would surpass
$\kcost[\partnbar,\bldkbar](z^n)$, which is $O(n)$. Hence, for
sufficiently large $n$, denoting by $\Perr(n)$ the probability of a
pair from $\sspace_2$ prevailing over $(\partnbar,\bldkbar)$
in~(\ref{eq:rulePi}), and observing that $\alpha^\alpha
(\log(n+1))^{\alpha+1}
\le (n+1)^{{\alpha\log\alpha + \alpha + 1}}$ for $n\ge 1$, it
follows from~(\ref{eq:Pbad}) that
\begin{eqnarray*}
\Perr(n) &\le&
\sum_{(\partnp,\bldk'):\dpartni'\ge\dpartni_0} \Pbadp \le \sum_{\dpartni'=\dpartni_0}^{n}
\alpha^\alpha \left(\log
(n+1)\right)^{\alpha+1}(n+1)^{\beta(\dpartnib-\dpartni') + \dpartni'}
\\
&\le&\sum_{\dpartni'=\dpartni_0}^{n}(n+1)^{\dpartni'(1-\beta)+\beta\dpartnib
+{\alpha\log\alpha+\alpha+1}}
\le(n+1)^{\dpartni_0(1-\beta)+\beta\dpartnib
+\alpha\log\alpha+\alpha+2},
\end{eqnarray*}
where the last inequality holds for $\beta > 1$. Choosing
$\dpartni_0>\frac{\beta\dpartnib+\alpha\log\alpha+\alpha+3}{\beta-1}$,
we get
\[
\Perr(n) \le (n+1)^{\delta}\,,
\]
for a constant $\delta < -1$. Therefore, $\Perr(n)$ is summable, and,
applying again Borel-Cantelli's lemma,
$(\hat{\partn},\hat{\bldk})\not\in\sspace_2$ a.s.\ as $n\to\infty$. We
conclude that $(\hat{\partn},\hat{\bldk})$ is compatible with $P$ a.s.\
as $n\to\infty$, as claimed. The fact that $\hat{\bldk}$ is, almost
surely, the correct order vector follows from the well known
consistency of penalized ML estimators for Markov
order~\cite{CsiszarShields00} (recall, from the discussion
following~(\ref{eq:rulePi}), that the order of each subprocess is
estimated independently).

 The second claim of the theorem is proved by applying
Lemma~\ref{lem:canonicalwins}, which implies that in the
\domfree\ case, the canonical partition beats other compatible
partitions with more subalphabets. When $k_\swltr{>}0$, this follows
from~(\ref{eq:aswin}), while when $k_\swltr{=}0$, it follows
from~(\ref{eq:tie}) and our tie-breaking convention.
\end{proof}

\medskip

\noindent{\textbf{Acknowledgment.}
Thanks to Erik Ordentlich and Farzad Parvaresh for stimulating
discussions.}

\medskip
\bibliographystyle{IEEEtran}
%Included for Gather Purpose only:
%input "memswitch.bib"
\bibliography{memswitch}
\end{document}